\documentclass[journal,draftcls,onecolumn]{IEEEtran}
\usepackage[centertags]{amsmath}
\usepackage{epsf,epsfig,cite,caption}
\usepackage{amsfonts}
\usepackage{amssymb}
\usepackage{dsfont,newlfont,latexsym,mathrsfs,url}
\newcommand{\calA}{{\cal A}}

\newcommand{\calC}{{\cal C}}

\newcommand{\calG}{{\cal G}}

\newcommand{\calM}{{\cal M}}

\newcommand{\calP}{{\cal P}}
\newcommand{\calQ}{{\cal Q}}

\newcommand{\calS}{{\cal S}}

\newcommand{\calU}{{\cal U}}
\newcommand{\calV}{{\cal V}}

\newcommand{\calX}{{\cal X}}
\newcommand{\calY}{{\cal Y}}

\newcommand{\bc}{{\mathbf c}}

\newcommand{\bk}{{\mathbf k}}
\newcommand{\bK}{{\mathbf K}}

\newcommand{\bs}{{\mathbf s}}
\newcommand{\bS}{{\mathbf S}}
\newcommand{\bu}{{\mathbf u}}

\newcommand{\bx}{{\mathbf x}}
\newcommand{\bX}{{\mathbf X}}
\newcommand{\by}{{\mathbf y}}
\newcommand{\bY}{{\mathbf Y}}
\newcommand{\bz}{{\mathbf z}}

\newcommand{\tp}{{\tilde p}}
\newcommand{\dsf}{{\mathsf d}}
\newcommand{\pYX}{p_{\bY|\bX}}
\newcommand{\dotle} {\mbox{$\:\stackrel{\centerdot}{\le}\:$}}
\newcommand{\dotge} {\mbox{$\:\stackrel{\centerdot}{\ge}\:$}}

\newtheorem{theorem}{Theorem}

\newtheorem{lemma}{Lemma}
\newtheorem{definition}{Definition}

\newtheorem{proposition}{Proposition}
\begin{document}
\title{Perfectly Secure Steganography: Capacity, Error Exponents, and Code Constructions}
\author{Ying~Wang,~\IEEEmembership{Student Member,~IEEE,}
and~Pierre~Moulin,~\IEEEmembership{Fellow,~IEEE}\thanks{
This work was supported by NSF under grants CCR 02-08809 and CCR
03-25924, and presented in part at the 40th Conference on
Information Sciences and Systems (CISS), Princeton, NJ, March
2004. Ying Wang was with the Department of Electrical and Computer
Engineering at the University of Illinois at Urbana-Champaign and
is now with Qualcomm Flarion Technologies, Bridgewater, NJ 08807
USA (e-mail: yingw@qualcomm.com). Pierre Moulin is with the
Beckman Institute, Coordinate Science Lab and Department of
Electrical and Computer Engineering, University of Illinois at
Urbana-Champaign, Urbana, IL 61801 USA (e-mail:
moulin@ifp.uiuc.edu).}}

\date{February 16, 2007. Revised September 30, 2007}
\maketitle

\begin{abstract}
An analysis of steganographic systems subject to the following
{\em perfect undetectability} condition is presented in this
paper. Following embedding of the message into the covertext, the
resulting stegotext is required to have {\em exactly} the same
probability distribution as the covertext. Then no statistical
test can reliably detect the presence of the hidden message. We
refer to such steganographic schemes as {\em perfectly secure}. A
few such schemes have been proposed in recent literature, but they
have vanishing rate. We prove that communication performance can
potentially be vastly improved; specifically, our basic setup assumes
independently and identically distributed (i.i.d.) covertext, and we
construct perfectly secure steganographic codes from public watermarking
codes using binning methods and randomized permutations of the code.
The permutation is a secret key shared between encoder and decoder.
We derive (positive) capacity and random-coding exponents for perfectly-secure
steganographic systems. The error exponents provide estimates of the code length
required to achieve a target low error probability.

In some applications, steganographic communication may be disrupted by
an {\em active warden}, modelled here by a compound discrete
memoryless channel. The transmitter and warden are subject to
distortion constraints. We address the potential loss in
communication performance due to the perfect-security requirement.
This loss is the same as the loss obtained under a weaker
{\em order-1} steganographic requirement that would just require
matching of first-order marginals of the covertext and stegotext
distributions. Furthermore, no loss occurs if the covertext distribution
is uniform and the distortion metric is cyclically symmetric;
steganographic capacity is then achieved by randomized linear
codes. Our framework may also be useful for developing computationally
secure steganographic systems that have near-optimal communication performance.
\end{abstract}

\IEEEpeerreviewmaketitle
\newpage

\section{Introduction}
\label{sec:introduction} Information embedding refers to the
embedding of data within a cover object (also referred to as
\emph{covertext}) such as image, video, audio, graphics, text, or
packet transmission times
\cite{Johnson_Katzenbeisser_2000,Johnson-JDJ00,
Cox02,Moulin_Koetter_IEEE05,Giles_Hajek_IEEEII03}. Applications
include copyright protection, database annotation, transaction
tracking, traitor tracing, timing channels, and multiuser
communications. These applications often impose the requirement
that embedding only slightly perturbs the covertext. The name {\em
watermarking} has been widely used to describe information
embedding techniques that are perceptually transparent, i.e., the
marked object (after embedding) is perceptually similar to the
cover object.

In some applications, the presence of the embedded information
should be kept secret (see applications below). Then perceptual
transparency is not sufficient, because statistical analysis could
reveal the presence of hidden information. The problem of
embedding information that is hard to detect is called
\emph{steganography}, and the marked object is called {\em
stegotext} \cite{Cox02,Moulin_Koetter_IEEE05,Anderson_Petitcolas_IEEEJSAC98,
Johnson-JJ98,Provos03}.
Steganography differs from cryptography in that the presence of
the message needs to remain secret, rather than the value of the
message. The dual problem to steganography is {\em steganalysis},
that is, detection of hidden information within a stegotext.

A classical model for steganography is Simmons' prisoner
problem~\cite{Simmons83}. Alice and Bob are locked up in different
cells but are allowed to communicate under the vigilant eye of
Willie, the prison warden. If Willie detects the presence of
hidden information in the transmitted data, he terminates their
communication and subjects them to a punishment. Willie is a {\em
passive warden} if he merely observes and analyzes the transmitted
data. He is an {\em active warden} if he introduces noise to make
Alice and Bob's task more difficult.

In the information age, there are several application scenarios
for steganography.
\begin{enumerate}
\item Steganography may be used to communicate over public networks such as
    the Internet. One may embed bits into inconspicuous files that
    are routinely sent over such networks: images, video, audio files, etc.
    Users of such technology may include intelligence and military personnel,
    people that are subject to censorship, and more generally,
    people who have a need for privacy.
\item Steganography may also be used to communicate over private networks.
    For instance, confidential documents within a commercial or governmental
    organization could be marked with identifiers that are hard to
    detect. The purpose is to trace unauthorized use of a document to a particular
    person who received a copy of this document. The recipient of the marked
    documents should not be aware of the presence of these identifiers.
\item Timing channels can be used to leak out information about computers.
    A pirate could modify the timing of packets sent by the computer,
    encoding data that reside on that computer. The pirate wishes to make
    this information leakage undetectable to avoid arousing suspicion.
    To disrupt potential information leakage, the network could jam packet timings
    --- hence the network plays the role of an active warden.
\end{enumerate}
The channel over which the stegotext is transmitted could be
noiseless or noisy, corresponding to the case of a passive and an
active warden, respectively. Moreover, the steganographer's
ability to choose the covertext is often limited if not altogether
nonexistent. In the private-network application above, the
covertext is generated by a content provider, not by the
steganographer (i.e., the authority responsible for document
security). Similarly in the timing-channel application, the
covertext is generated by the computer, not by the pirate.

In view of these applications, the four basic attributes of a
steganographic code are:
\begin{enumerate}
\item {\bf detectability:} quantifying Willie's ability to detect
    the presence of hidden information;
\item {\bf transparency (fidelity):} closeness of covertext and stegotext under
    an appropriate distortion (fidelity) metric;
\item {\bf payload:} the number of bits embedded in the covertext;
and \item {\bf robustness:} quantifying decoding reliability
    in presence of channel noise (i.e., when Willie is an active warden).
\end{enumerate}
If Alice had complete freedom for choosing the covertext,
the transparency requirement would be immaterial.
A covertext would not even be needed: it would suffice for Alice
to generate objects that follow a prescribed covertext
distribution. This model has two shortcomings: (a) as mentioned above,
in some applications Alice has little or no control over the choice
of the covertext; (b) even if she has, covertexts have complicated distributions,
and generating a size-$M$ steganographic code by sampling the covertext
distribution would be highly impractical for large $M$.

Information theory is a natural framework for studying steganography and steganalysis.
Assuming a statistical model is available
for covertexts, the only truly secure strategy from the steganographer's
point of view is to ensure that the probability distributions of
the covertext and stegotext are {\em identical}. This strong notion
of security was proposed by Cachin \cite{Cachin} and is the steganographic
counterpart of Shannon's notion of perfect security in cryptography.
We refer to steganography that satisfies this strong property
as \emph{perfectly secure}.

If Alice is allowed to select the covertext and Willie is passive,
Alice may use the following perfectly secure steganographic code \cite{Cachin}.
Alice and Bob agree on a hash function, and the value of the hashed stegotext
is the message to be transmitted.
Alice searches a database of covertexts until she finds one that
matches the desired hash value. This approach is perfectly secure
irrespective of the distribution of the covertext. The disadvantages are
that the search is computationally infeasible
for large message sets (communication rate is extremely low),
and the underlying communication model is limited, as discussed above.

Cachin also proposed two less stringent requirements for steganographic codes
~\cite{Cachin}. One is $\epsilon$-secure steganographic codes, where the
Kullback-Leibler divergence between the covertext and stegotext
probability distributions is smaller than $\epsilon$
(perfect security requires $\epsilon=0$). For random processes
he redefined perfectly secure steganography by requiring
that the above Kullback-Leibler divergence, normalized by the length $N$
of the covertext sequence, tends to zero as $N \to \infty$.
Unfortunately this does not preclude the possibility that Kullback-Leibler
divergence remains bounded away from zero, even grows to infinity
(at a rate slower than $N$) as $N \to \infty$. If such is
the case, Willie's error probability tends to zero asymptotically,
and therefore the perfect-security terminology is misleading.

While Cachin focused on security and not on communication
performance in terms of payload, robustness and fidelity,
Kullback-Leibler divergence has become a popular metric
for assessing the security of practical steganographic schemes
subject to transparency, payload, and robustness requirements
\cite{Wang_Moulin_SSP03, Wang_Moulin_SPIE04, Dabeer_etal_IEEESP04,
Moulin_Briassouli_ICIP04, Wang_Moulin_IFS2007,
Sullivan_steganalysis_IFS06, Solanki_zeroKL_ICIP06,
Sullivan_rate_ICIP06}. Other metrics are studied in
\cite{Harmsen_Pearlman_SPIE03,Lyu_Farid_IEEEIFS_2006,
Goljan_Fridrich_Holotyak_SPIE06,
Farid_ICIP02,
Fridrich_Goljan_Du_IEEEMM01,
Fridrich_Goljan_SPIE02, Dumitrescu_Wu_Wang_IEEESP03}.

The tradeoffs between detectability, fidelity, payload, and
robustness can be studied in an information-theoretic framework.
The basic mathematical model for steganography is communications
with side information at the encoder~\cite{Gel'fand_Pinsker_1980}.
Moulin and O'Sullivan studied a general information-theoretic
framework for information hiding and indicated its applicability
to steganography~\cite[Section~VII.C]{Moulin_O'Sullivan_IEEE03}.
However, they did not study perfectly secure steganography and did
not derive expressions for steganographic capacity. Galand and
Kabatiansky~\cite{Galand03} constructed steganographic binary
codes, but the code rate vanishes as $\frac{\log N}{N}$. Fridrich
{\em et
al.}~\cite{Fridrich_Goljan_Lisonek_Soukal_wetpaper_IEEESP05,
Fridrich_Goljan_Soukal_wetpaper_IEEEIFS06} proposed positive-rate
``wet paper'' codes, which permit a change from the original cover
distribution to a new stegodistribution. However they did not
analyze the fundamental tradeoffs between payload, robustness, and
detectability.

The goal of this paper is therefore to study the
information-theoretic limits of {\em perfectly undetectable}
steganography. As a first step towards this problem, we assume
that covertext samples are independently and identically
distributed (i.i.d.) over a finite alphabet. In practice the
i.i.d. model could be applied to transform coefficients or to
blocks of coefficients. While this is just a simplifying
approximation to actual statistics, it allows us to derive
tangible mathematical results and to understand the effects of
the perfect security constraint on transparency, payload,
and robustness.
Our first result is a connection between public watermarking
codes~\cite{Moulin_O'Sullivan_IEEE03,Somekh-Baruch_Merhav_PriWat2003,
Somekh-Baruch_Merhav_PubWat2004} and perfectly secure
steganographic codes. Given any public watermarking code that
preserves the {\em first-order statistics} of the covertext (this
property will be referred to as {\em order-1 security}), we show
that a perfectly secure steganographic code with the same error
probability can be constructed using randomization over the set of
all permutations of $\{ 1, 2, \cdots, N\}$. We use this construction to
derive capacity and random-coding exponent formulas for perfectly
secure steganography.

The codes that achieve capacity and random-coding exponents are
stacked-binning schemes as proposed in~\cite{Moulin_Wang_IEEEIT04}
for general problems of channel coding with side information. The
random-coding exponent yields an asymptotic upper bound on
achievable error probability and therefore serves as an estimate
of the code length required to achieve a target low error
probability. A stacked-binning code consists of a stack of
variable-size codeword arrays indexed by the type of the covertext
sequence, and the corresponding decoder is a maximum penalized
mutual information (MPMI) decoder.  The analysis is based on the
method of types \cite{Csiszar_Korner_book1997,Csiszar98}.

Due to the added perfect-security constraint, capacity and random-coding
exponent for steganography cannot exceed those of the
corresponding public watermarking problem. Nevertheless, we have
identified a class of problems where the covertext probability
mass function (PMF) is uniform and the distortion function is
symmetric, with the property that the perfect undetectability
constraint does not cause any capacity loss. One special example
in the general class is the case of Bernoulli($\frac{1}{2}$)
covertexts with the Hamming distortion
metric~\cite{Moulin_Wang_CISS04}. For the binary-Hamming case, the
perfect security condition has no effect on both the capacity and
random-coding error exponent. Steganographic capacity is achieved
by randomized nested linear codes.

This paper is organized as follows. Section~\ref{sec:notation}
describes the notation, and Section~\ref{sec:setup} the problem
statement. Section~\ref{sec:codes} shows how perfectly secure
steganographic codes can be constructed from codes with the much
weaker order-1 security. Section~\ref{sec:Stegcap main results} presents
our main theorems on capacity and random-coding error exponent.
Section~\ref{sec:secret key} discusses the role of secret keys in
steganographic codes. Simplified results for the no-attack case
are stated in Section~\ref{sec:passive}. A class of steganography
problems for which perfect security comes at no cost is studied in
Section~\ref{sec:loss}. As an example of this class, the
binary-Hamming problem is studied in
Section~\ref{sec:binary-hamming}.
The paper concludes with a discussion
in Section~\ref{sec:Steg Discussion}.

\section{Notation}
\label{sec:notation}
We use uppercase letters for random variables, lowercase letters
for their individual values, and boldface letters for sequences.
The PMF of a random variable $X \in \calX$ is denoted by
$p_X=\{p_X(x),\,x \in \calX\}$. The entropy of a random variable
$X$ is denoted by $H(X)$, and the mutual information between two
random variables $X$ and $Y$ is denoted by $I(X;Y)=H(X)-H(X|Y)$.
Should the dependency on the underlying PMFs be explicit, we use
the PMFs as subscripts, e.g., $H_{p_X}(X)$ and
$I_{p_X,p_{Y|X}}(X;Y)$. The Kullback-Leibler divergence between
two PMFs $p$ and $q$ is denoted by $D(p||q)$; the conditional
Kullback-Leibler divergence of $p_{Y|X}$ and $q_{Y|X}$ given $p_X$
is denoted by
$D(p_{Y|X}||q_{Y|X}|p_X)=D(p_{Y|X}\,p_X||q_{Y|X}\,p_X)$.

Let $p_\bx$ denote the empirical PMF on $\calX$ induced by a
sequence $\bx \in \calX^N$. Then $p_\bx$ is called the type of
$\bx$. The type class $T_\bx$ associated with $p_\bx$ is the set
of all sequences of type $p_\bx$. Likewise, we define the joint
type $p_{\bx\by}$ of a pair of sequences $(\bx, \by) \in \calX^N
\times \calY^N$ and the type class $T_{\bx\by}$ associated with
$p_{\bx\by}$.
The conditional type $p_{\by|\bx}$ of a pair of sequences ($\bx, \by$)
is defined as $\frac{p_{\bx\by}(x,y)}{p_{\bx}(x)}$ for
all $x \in \calX$ such that $p_{\bx}(x) > 0$. The conditional type
class $T_{\by|\bx}$ given $\bx$ is the set of all sequences $\tilde{\by}$
such that $(\bx, \tilde{\by}) \in T_{\bx\by}$. We denote by $H(\bx)$
the empirical entropy for $\bx$, i.e.,
the entropy of the empirical PMF $p_{\bx}$.
Similarly, we denote by $I(\bx;\by)$ the empirical
mutual information for the joint PMF $p_{\bx\by}$. The above
notation for types is adopted from Csisz\'{a}r and
K\"{o}rner~\cite{Csiszar_Korner_book1997}.

We let $\mathbb U(\Omega)$ denote the uniform PMF over a finite set
$\Omega$. We let $\calP_X$ and $\calP^N_X$ represent the set of
all PMFs and all empirical PMFs, respectively, on the alphabet
$\calX$. Likewise, $\calP_{Y|X}$ and $\calP^N_{Y|X}$ denote the
set of all conditional PMFs and all empirical conditional PMFs on
the alphabet $\calY$. We use $\mathsf E$ to denote
mathematical expectation.

The shorthands $a_N \doteq b_N$, $a_N \dotle b_N$, and $a_N \dotge
b_N$ are used to denote asymptotic equalities and inequalities in
the exponential scale for $\lim_{N \to \infty}\frac{1}{N}\log
\frac{a_N}{b_N}=0$, $\limsup_{N \to \infty}\frac{1}{N}\log
\frac{a_N}{b_N}\le 0$, and $\liminf_{N \to \infty}\frac{1}{N}\log
\frac{a_N}{b_N}\ge 0$, respectively. We define $|t|^+ \triangleq
\max(t,0)$, $\exp_2(t)\triangleq 2^t$, and the binary entropy
function
\[h(t)\triangleq -t\log t -(1-t) \log (1-t),\quad  t\in [0, 1].\]
We use $\ln x$ to denote the natural logarithm of $x$, and the
logarithm $\log x$ is in base 2 if not specified otherwise. The
notation $\mathds1_{\{A\}}$ is the indicator function of the
event $A$:
\begin{equation}
\mathds1_{\{A\}}=\left\{\begin{array}{ll}1& A \mbox{ is
true;}\\0& \mbox{ else.}\end{array}\right.\nonumber
\end{equation}
Finally, we adopt the notional convention that the minimum (resp.
maximum) of a function over an empty set is $+\infty$ (resp. 0).
\section{Problem Statement}
\label{sec:setup}
\begin{figure}[hbt]
   \begin{center}
   \begin{tabular}{c}
  \hspace{-0.5cm} \includegraphics[height=4cm]{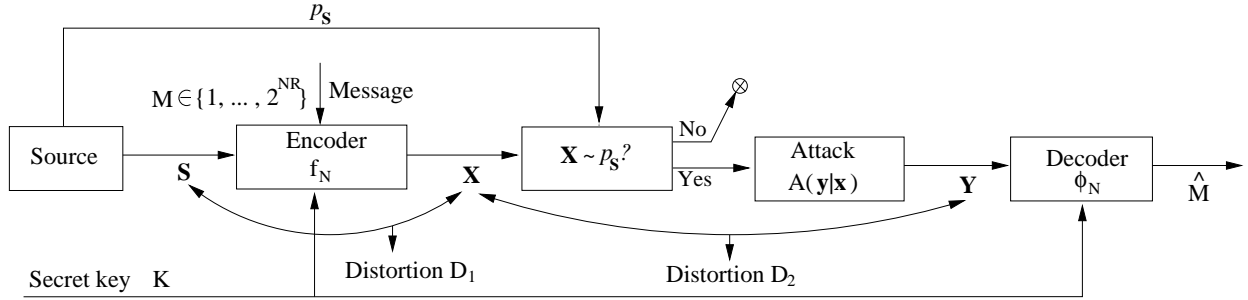}
   \end{tabular}
   \end{center}
   \caption{Communication-theoretic view of perfectly secure steganography.}
    \label{fig:steg}
   \end{figure}
Referring to Fig.~\ref{fig:steg}, the covertext is modelled as a
sequence $\bS=(S_1, \cdots, S_N)$ of i.i.d. samples drawn from a
PMF $\{ p_S(s)$, $s \in \calS \}$. A message $M$ is to be embedded
in $\bS$ and transmitted to a decoder; $M$ is uniformly
distributed over a message set $\calM$. The encoder produces a
stegotext $\bX$ through a function $f_N(\bS, M)$, in an attempt to
transmit the message $M$ to the decoder reliably. The covertext
and stegotext are required to be close according to some
distortion metric. The distortion model is motivated by the fact that
stegotext and covertext represent physical signals (such as images,
text, etc.) which can be modified to some extent without affecting
perceptual quality \cite{Moulin_O'Sullivan_IEEE03}.
The strength of the transparency constraint is controlled by
a distortion parameter.

A steganalyzer observes $\bX$ and tests whether $\bX$ is drawn
i.i.d. from $p_S$. If not, Willie, the steganalyzer terminates the
transmission, and obviously the decoder is unable to retrieve $M$.
If $\bX$ is deemed innocuous, Willy may simply forward it to the
decoder. In this case, Willie is a passive warden. To be on the safe side
for preventing reliable transmission of hidden messages, Willie may
want to pass $\bX$ through some attack channel $\pYX(\by|\bx)$, thereby
producing a corrupted text $\bY$. In this case, Willie is an active warden,
and the corrupted text and the stegotext are also required to be close according
to some distortion metric. The alphabets $\calS$, $\calX$ and $\calY$
for $S$, $X$ and $Y$, respectively, are assumed to be identical.

The decoder does not know $\pYX$ selected by the steganalyzer and
does not have access to the original covertext $\bS$. The decoder
produces an estimate $\hat{M} = \phi_N(\bY) \in \calM$ of the
transmitted message. We assume that the encoder/decoder pair
$(f_N, \phi_N)$ is {\em randomized}, i.e., the choice of $(f_N,
\phi_N)$ is a function of a random variable known only to the
encoder and decoder but not to the steganalyzer. We can think of
this random variable as a {\em secret key} as
in~\cite{Moulin_O'Sullivan_IEEE03,Somekh-Baruch_Merhav_PriWat2003,
Somekh-Baruch_Merhav_PubWat2004}.
Note that in generic information-hiding games, this secret key
provides some protection against adversaries with arbitrary memory
and unlimited computational
resources~\cite[Section~X]{Moulin_Koetter_IEEE05}. In steganography,
the secret key plays a fundamental role in ensuring
perfect undetectability: the covertext and the stegotext have the same
PMF when the secret key is carefully designed. The randomized code
will be denoted by $(F_N, \Phi_N)$ with a joint distribution
$p(f_N,\phi_N)$.
%
%

\subsection{Steganographic Codes}
\label{sec:Constrained Steganographic Codes}
A distortion function
is any nonnegative function $\dsf:\calS \times \calS \to \mathbb
R^+ \cup \{0\}$. This definition is extended to length-$N$ vectors
using $\dsf^N(\bs,\bx)=\frac{1}{N}\sum_{i=1}^N \dsf(s_i,x_i)$. Let
$D_{\max} = \max_{s,x} \dsf(s,x)$. We assume without loss of
generality that $\dsf(s,x) \ge 0$, with equality if $s=x$.
\bigskip
\begin{definition}
A length-$N$ {\bf perfectly secure steganographic code} with
maximum distortion $D_1$ is a triple $(\calM, F_N, \Phi_N)$, where
\begin{itemize}
\item $\calM$ is the message set of cardinality $|\calM|$;
\item $(F_N, \Phi_N)$ has a joint distribution $p(f_N,\phi_N)$;
\item $f_N~:~\calS^N \times \calM \to \calS^N$ maps covertext $\bs$ and
    message $m$ to stegotext $\bx=f_N(\bs,m)$. The mapping is subject
    to the maximum distortion constraint
\begin{equation}
\dsf^N(\bs,f_N(\bs,m)) \le D_1 \mbox{ almost surely }
\label{eq:D1}
\end{equation}
and the {\bf perfect undetectability} constraint
\begin{equation}
   p_{\bX} = p_{\bS};
\label{eq:secure}
\end{equation}
\item $\phi_N~:~\calS^N \to \calM$ maps 
    the received sequence $\by$ to a decoded message
    $\hat m=\phi_N(\by)$.
\end{itemize}
\label{def:PS codes}
\end{definition}

The above definition is similar to the definitions for a
length-$N$ data-embedding or watermarking code
in~\cite{Moulin_O'Sullivan_IEEE03,Somekh-Baruch_Merhav_PriWat2003,
Somekh-Baruch_Merhav_PubWat2004}, with the additional steganographic constraint
of (\ref{eq:secure}) which requires perfect matching of $N$-dimensional
distributions. Also observe that the distortion constraint is inactive if
$D_1 \ge D_{\max}$, i.e., the covertext $\bS$ available to Alice plays no role.
Given $p_S$, define the set of conditional PMFs $p_{X|S}$ such that the marginals
of $p_S p_{X|S}$ are equal ($p_X = p_S$) and the expected distortion between
$S$ and $X$ does not exceed $D_1$:
\begin{equation}
   \calQ_1^{Steg}(p_S,D_1) \triangleq \left\{ p_{X|S} ~:~
        \sum_{s,x} p_{X|S}(x|s) \,p_S(s)\, \dsf(s,x) \le D_1 , \quad
        p_X(x) = \sum_s p_{X|S}(x|s)\, p_S(s) = p_S(x), \, \forall x \in \calS \right\} .
\label{eq:Q1steg}
\end{equation}
Also recall that in Def.~\ref{def:PS codes}, randomization of $(F_N, \Phi_N)$
is realized via a cryptographic key shared by encoder and decoder.

Next, we define CCC and RM codes which will be used
to construct perfectly secure steganographic codes.
\begin{definition}
(CCC Code).
A length-$N$ code with {\bf conditionally constant composition},
{\bf order-1 steganographic property}, and {\bf maximum distortion} $D_1$
is a quadruple $(\calM, \Lambda, F_N, \Phi_N)$,
where $\Lambda$ is a mapping from $\calP_S^{[N]}$ to $\calP_{X|S}^{[N]}$.
The transmitted sequence $\bx = f_N(\bs,m)$ has conditional type
$p_{\bx|\bs} = \Lambda(p_{\bs})$.
Moreover, $\Lambda(p_{\bs}) \in \calQ_1^{Steg}(p_{\bs},D_1)$.
\label{def:CCC codes}
\end{definition}

Observe that such a code matches the first-order empirical
marginal PMF of the covertext, but not necessarily higher-order
empirical marginals. Hence such a code generally does not  satisfy
the perfect-undetectability property.
\begin{definition}
(RM Code).
A length-$N$ {\bf randomly modulated} code is the randomized code defined
via permutations of a prototype ($f_N,\phi_N$):
\begin{eqnarray}
\bx=f_N^\pi(\bs,m) & \triangleq & \pi^{-1}f_N(\pi \bs,m)  \label{eq:rm-f} \\
\phi_N^\pi(\by) & \triangleq & \phi_N(\pi \by), \label{eq:rm-phi}
\end{eqnarray}
where $\pi$ is drawn uniformly from the set $\Pi$ of all $N!$
permutations and is not revealed to Willie. The sequence
$\pi \bx$ is obtained by applying $\pi$ to the elements of $\bx$.
\label{def:RM codes}
\end{definition}
\begin{definition}
Given alphabets $\calS$ and $\calU$, a \textbf{steganographic channel}
$p_{XU|S}(x,u|s)$ subject to distortion $D_1$ is a conditional PMF
whose conditional marginal $p_{X|S}$ belongs to $\calQ_1^{Steg}(p_S,D_1)$
of (\ref{eq:Q1steg}).
We denote by $\calQ^{Steg}(L,p_S,D_1)$ the set of steganographic channels
subject to distortion $D_1$ when the alphabet $\calU$ has cardinality $L$.
\label{def:covert channel}
\end{definition}

If the channel $p_{XU|S}$ satisfies the distortion constraint $D_1$
but not necessarily the steganographic constraint $p_X = p_S$,
$p_{XU|S}$ is simply a covert channel in the sense
of \cite{Moulin_O'Sullivan_IEEE03,Somekh-Baruch_Merhav_PriWat2003}.
We shall denote by $\calQ (L,p_S,D_1)$ the set of all such covert
channels. Clearly, $\calQ^{Steg}(L,p_S,D_1) \subseteq \calQ (L,p_S,D_1)$.

\subsection{Attack Channels}
\label{sec:Constrained Attack Channels}
A passive warden simply produces $\bY=\bX$.
An active warden passes $\bX$ through a discrete memoryless channel
(DMC), producing a degraded sequence $\bY$.
\begin{definition}
A discrete memoryless attack channel $p_{Y|X}$ is feasible if
the expected distortion between $X$ and $Y$ is at most $D_2$:
\begin{equation}
   \sum_{x,y} p_X(x)\, p_{Y|X}(y|x)\, \dsf(x,y) \le D_2.
\label{eq:DMC-D2}
\end{equation}
Then the joint conditional PMF is given by \[\pYX(\by|\bx) =
\prod_{i=1}^N p_{Y|X}(y_i|x_i).\] We denote by
\[ \calA(p_X,D_2) = \left\{ p_{Y|X} \in \calP_{Y|X} ~:~
    \sum_{x,y} p_X(x)\, p_{Y|X}(y|x)\, \dsf(x,y) \le D_2 \right\} \]
the set of all such feasible DMCs. This set is a compound
DMC family. \label{def:DMC}
\end{definition}

As an alternative to Def.~\ref{def:DMC}, one may consider
attack channels that have arbitrary memory but are subject to an
almost sure distortion constraint~\cite{Somekh-Baruch_Merhav_PriWat2003,
Somekh-Baruch_Merhav_PubWat2004,Moulin_Wang_IEEEIT04}.
In this case, the set of feasible attack channels is given by
\[\calA^\prime(p_\bx,D_2) = \left\{p_{\bY|\bX} \in \calP^N_{Y|X}:
    Pr\left[\dsf^N(\by,\bx) \le D_2\right]=1 \right\} .\]
There are three reasons why only memoryless channels are
considered in this paper. First, it is shown
in~\cite{Moulin_Wang_IEEEIT04} that for watermarking problems,
both DMCs with expected distortion and arbitrary memory attack
channels with almost sure distortion result in the same capacity
formula, and the former allows a smaller random-coding error
exponent when $D_2$ is the same. Thus, in terms of minimizing the
random-coding exponent, selecting $p_{Y|X}$ from the compound DMC
class $\calA(p_X,D_2)$ is a better strategy for the warden than
selecting $p_{\bY|\bX}$ from $\calA^\prime(p_\bx,D_2)$. Second,
the assumption of memorylessness simplifies the presentation of
main ideas. Finally, note that the proofs for the compound DMC
provide the  basis for the proofs in the case of channels with
arbitrary memory
\cite{Somekh-Baruch_Merhav_PubWat2004,Moulin_Wang_IEEEIT04}.

\subsection{Steganographic Capacity and Reliability Function}
\label{sec:Capacity and Reliable Function}
The probability of error for a randomized code ($F_N$,
$\Phi_N$) under a channel $p_{Y|X}$ is given by
\begin{eqnarray}
P_{e,N}(F_N, \Phi_N, p_{Y|X})=Pr(\hat M \neq M),
\end{eqnarray}
where the average is over all possible covertexts $\bS$, messages $M$,
and codes ($F_N$, $\Phi_N$).
\begin{definition}
A rate $R$ is achievable if there exists a randomized code $(F_N,
\Phi_N)$ such that  $|\calM| \ge 2^{NR}$ and
\begin{equation}
\sup_{p_{Y|X}} P_{e,N}(F_N, \Phi_N, p_{Y|X}) \to 0
    \quad \mbox{ as } N \to \infty.
\end{equation} \label{def:rate}
\end{definition}
\begin{definition}
The steganographic capacity $C^{Steg}(D_1, D_2)$ is the supremum
of all achievable rates. \label{def:capacity}
\end{definition}
\begin{definition}
The steganographic reliability function $E^{Steg}(R)$ is defined
as
\begin{equation}
E^{Steg}(R)=\liminf_{N \to \infty}\left[-\frac{1}{N}\log
   \inf_{F_N, \Phi_N} \sup_{p_{Y|X}} P_{e,N}(F_N, \Phi_N, p_{Y|X}) \right].
\end{equation}
\label{def:reliability}
\end{definition}

\section{From Order-1 to Perfectly Secure Steganographic Codes}
\label{sec:codes}
Codes with conditionally constant composition (Def.~\ref{def:CCC codes})
and randomly modulated codes (Def.~\ref{def:RM codes}) play a central
role in our code constructions and coding theorems.
The following proposition suggests a general construction for perfectly secure
steganographic codes: first select some deterministic prototype $f_N$ with the CCC and
order-1 steganographic properties and maximum distortion $D_1$ (Def.~\ref{def:CCC codes}),
second construct a RM code from that prototype.
In Section~\ref{sec:Stegcap main results} we show that this strategy
is an optimal one. The proof of the proposition appears in the appendix.
\begin{proposition}
Let $(\calM, F_N, \Phi_N)$ be a RM code whose prototype
$(f_N, \phi_N)$ has conditionally constant composition, order-1 security,
and maximum distortion $D_1$.
Then $(\calM, F_N, \Phi_N)$ is a perfectly secure steganographic code
with maximum distortion $D_1$ and same error probability as the prototype
($f_N, \phi_N$).
\label{prop:order1}
\end{proposition}

\section{Steganographic Capacity and Random Coding Error Exponent}
\label{sec:Stegcap main results}
The steganographic codes in our achievability proofs are randomly-modulated
binning codes with conditionally constant composition. The existence of
a good deterministic prototype is established using a random coding argument.
An arbitrarily large integer $L$ is selected, defining
an alphabet $\calU = \{ 1, 2, \cdots, L\}$ for the auxiliary random variable $U$
in the binning construction.  Given the covertext $\bs$ and the message $m$,
the encoder selects an appropriate sequence $\bu$ in the binning code and
then generates the stegotext randomly according to the uniform distribution
over an optimized type class $T_{\bx|\bu,\bs}$. Proofs of
the theorem and propositions in this section appear in
Appendices~\ref{app:Stegconverse}-\ref{app:proposition 2 Steg}.

The following difference between two mutual informations:
\begin{equation}
J_L(p_S,p_{XU|S},p_{Y|XUS})\triangleq I(U;Y)-I(U;S)
\label{eq:J-L}
\end{equation}
plays a fundamental role in the analysis.
\begin{theorem}
Under Def.~\ref{def:PS codes} for steganographic codes and
Def.~\ref{def:DMC} for the compound attack channel, steganographic capacity
is given by
\begin{eqnarray}
C^{Steg}(D_1, D_2) &=& \lim_{L\to\infty} C_L^{Steg}(D_1,D_2),
\label{eq:C-steg}
\end{eqnarray}
where
\begin{eqnarray}
 C_L^{Steg}(D_1,D_2)\triangleq \max_{p_{XU|S} \in \calQ^{Steg}(L,p_S,D_1)} \min_{p_{Y|X}
\in \calA(p_X,D_2)} J_L(p_S,p_{XU|S},p_{Y|X}) \label{eq:C-L-steg}
\end{eqnarray}
and $(U,S) \to X \to Y$ forms a Markov chain.
\label{theorem:Stegcapacity}
\end{theorem}

The proof of Theorem~\ref{theorem:Stegcapacity} is given in two
parts. The converse part is proved in
Appendix~\ref{app:Stegconverse}. The direct part is a corollary of
a stronger result stated in Proposition~\ref{proposition:exponent}
below, which provides a lower bound on the achievable error
exponent (hence an upper bound on the average probability of
error) and is proved in Appendix~\ref{app:Stegdirect}.
\begin{proposition}
Under Def.~\ref{def:PS codes} for steganographic codes and
Def.~\ref{def:DMC} for the compound attack channel, the following
random-coding error exponent is achievable:
\begin{equation}
E_r^{Steg}(R)=\lim_{L \to \infty} E_{r,L}^{Steg}(R),
\label{eq:Er}
\end{equation}
where
\begin{eqnarray}
E_{r,L}^{Steg}(R)&\triangleq&\min_{\tilde p_S \in \mathcal
P_S} \max_{p_{XU|S} \in \calQ^{Steg}(L,\tilde p_S, D_1)}\min_{\tilde p_{Y|XUS}
\in \calP_{Y|XUS} }\min_{p_{Y|X} \in
\calA(p_X,D_2)} \label{eq:ErL}\\
&&\left[D(\tilde p_S\, p_{XU|S}\, \tilde p_{Y|XUS}||p_S\,
p_{XU|S}\, p_{Y|X})+\left|J_L(\tilde p_S, p_{XU|S}, \tilde
p_{Y|XUS})-R\right|^+\right].\nonumber
\end{eqnarray}
Moreover, $E_r^{Steg}(R)=0$ if and only if $R \ge C^{Steg}$.
\label{proposition:exponent}
\end{proposition}

\emph{Remark 1}: The capacity and error exponent formulas in
(\ref{eq:C-steg})---(\ref{eq:ErL}) coincide with those for public
watermarking~\cite{Moulin_Wang_IEEEIT04,
Somekh-Baruch_Merhav_PubWat2004}, the only difference being
that here the maximization over $p_{XU|S}$ is subject to
a steganographic constraint.
Clearly $E_{r,L}^{Steg}(R) \le E_{r,L}^{PubWM}(R)$
and $C^{Steg} \le C^{PubWM}$.

\emph{Remark 2}: Proposition~\ref{proposition:exponent} is proved using
a random binning technique. First we establish
the existence of a deterministic prototype CCC code with order-1
steganographic property, maximum distortion $D_1$, and error
exponent $E^{Steg}(R)$. The decoder is an MPMI decoder. The main
steps in this part of the proof are similar to those in the proof
of Theorem~3.2 in \cite{Moulin_Wang_IEEEIT04}, with the additional
order-1 steganographic constraint on the encoder. The second part
of the proof is an application of Proposition~\ref{prop:order1}:
random modulation of the CCC prototype code yields a perfectly-secure
steganographic code with maximum distortion $D_1$ and error exponent
$E^{Steg}(R)$.

\emph{Remark~3}. As mentioned earlier, the covertext plays no role
in the special case $D_1 \ge D_{\max}$, and so Alice can generate
$\bX$ independently of $\bS$. The capacity formula
(\ref{eq:C-steg}) becomes simply
\[ C^{Steg} = \min_{p_{Y|S} \in \calA(p_S,D_2)} I(S;Y) , \]
and the random-coding exponent is
\[ E_r^{Steg}(R) = \min_{\tp_S} \min_{\tp_{Y|S} \in \calP_{Y|S}}
    \min_{p_{Y|S} \in \calA(p_S,D_2)} \left[ D(\tp_{Y|S} \,\tp_S \| p_{Y|S} \,p_S)
        + |I_{\tp_S \tp_{Y|S}}(S;Y) - R|^+ \right] . \]
The binning codes are degenerate in this case; the expressions for
capacity and random-coding exponents reduce to classical formulas
for compound DMCs without side information \cite{Csiszar_Korner_book1997}
and are achieved using constant-composition codes.
Further specializing this result to the case of a passive warden
($D_2 = 0$, hence $p_{Y|X} = \mathds1_{\{Y=X\} })$, we obtain
$C^{Steg} = H(S)$ and $E_r^{Steg}(R)$ is given by (\ref{eq:E-steg-passive}),
see Section~\ref{sec:passive}.

\begin{figure}
   \begin{center}
   \begin{tabular}{c}
   \includegraphics[height=5cm]{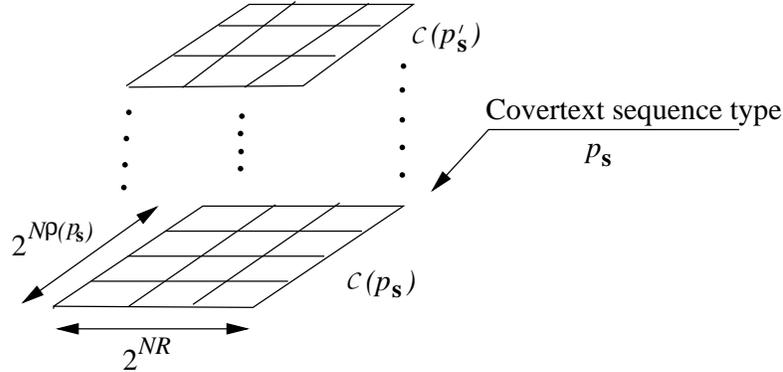}
   \end{tabular}
   \end{center}
   \caption{A binning scheme with a stack of variable-size codeword arrays
indexed by the covertext sequence type.}
 \label{fig:BinStack}
   \end{figure}

The operation of the deterministic prototype code is illustrated in
Fig.~\ref{fig:BinStack}. The codebook $\calC$ consists of a stack
of codeword arrays indexed by the possible covertext sequence
types. Given an input $\bs$, the encoder evaluates its type
$p_{\bs}$ and selects the corresponding codeword array
\begin{equation}
\calC(p_\bs)=\{\bu(l,m,p_\bs),\; 1\le l \le
2^{N\rho(p_\bs)},\;1\le m \le |\calM|\},
\end{equation}
in which the codewords are drawn from an optimized type class
$T_\bu \triangleq T_U^*(p_\bs)$. Each array $\calC(p_\bs)$
has $|\calM|$ columns and $2^{N \rho(p_\bs)}$ rows,
where $\rho(p_\bs)$ is a function of the corresponding covertext
type $p_\bs$ and is termed the depth parameter of the array.
Given $\by$, the decoder seeks a codeword in
$\calC =\bigcup_{p_\bs}\calC(p_\bs)$ that maximizes the penalized
empirical mutual information and outputs its column index as the
estimated message:
\begin{equation}
\hat m=\arg \max_m
\max_{l,p_\bs}\left[I(\bu(l,m,p_\bs);\by)-\rho(p_\bs)\right].
\end{equation}
By letting $\rho(p_\bs) = I(\bu;\bs)+\epsilon$, where $T_{\bu\bs}
\triangleq T_{US}^*(p_\bs)$ is an optimized joint type and
$\epsilon$ is an arbitrarily small positive number, an optimal
balance between the probability of encoding error and the
probability of decoding error is achieved. The former vanishes
double-exponentially while the latter vanishes at a rate given by
the random coding error exponent in (\ref{eq:ErL}). The above MPMI
decoder can be thought of as an empirical generalized maximum a
posterior (MAP) decoder~\cite[Section~3.1]{Moulin_Wang_IEEEIT04}.

\section{Secret Key}
\label{sec:secret key}
In standard information-hiding problems
with a compound DMC attack channel, \emph{deterministic} codes are
enough to achieve capacity; random coding is used as a method of
proof to establish the existence of a deterministic code without
actually specifying the
code~\cite{Lapidoth_ChannelUncertainty_1998}. In our steganography
problem, a {\em randomized} code is used to satisfy the
perfect-undetectability condition of (\ref{eq:secure}). Without
the secret key, a deterministic code generally could not satisfy
the perfect-undetectability condition. Also note that a randomized
code is generally needed if the attacks have arbitrary memory
\cite{Somekh-Baruch_Merhav_PriWat2003,
Somekh-Baruch_Merhav_PubWat2004, Moulin_Wang_IEEEIT04}. For
example, in watermarking games, knowing a deterministic code the
adversary would decode and remove the message; deterministic codes
are vulnerable to this kind of ``surgical
attack''~\cite{Moulin_Koetter_IEEE05}.

For randomized codes, the secret key shared between encoder and
decoder is the source of common randomness. For RM codes, the
secret key specifies the value of the permutation $\pi$. The
entropy rate of the secret key is
\begin{eqnarray}
H_K^{RM}=\frac{1}{N}\log_2 N!<\log_2 N. \label{eq:entropy rate E RM}
\end{eqnarray}

\section{Passive Warden}
\label{sec:passive}
A passive warden introduces no degradation to the stegotext;
in this case, $D_2=0$ and $Y = X$, i.e.,
\begin{equation}
   p_{Y|X} = \mathds1_{\{Y=X\} } .
\label{eq:passive}
\end{equation}
This results in simplified expressions for the perfectly secure
steganographic capacity in (\ref{eq:C-steg}) and the random-coding
error exponent in (\ref{eq:Er}), see
Propositions~\ref{proposition:passive capacity}
and~\ref{proposition:passive error exponent} below.
The proofs of these propositions appear in Appendices
\ref{app:C-passive} and \ref{app:E-passive}, respectively.
\begin{proposition}
For the passive-warden case ($D_2=0$), the maximization in
(\ref{eq:C-L-steg}) is achieved by $U=X$ and
\begin{equation}
   C^{Steg}(D_1, 0) = \max_{p_{X|S} \in \calQ_1^{Steg}(p_S,D_1)} H(X|S) .
\label{eq:C-steg-passive}
\end{equation}
\label{proposition:passive capacity}
\end{proposition}

\emph{Remark}.
Since $H(X|S) = H(X) - I(S;X) = H(S) - I(S;X)$, we have
\[ C^{Steg}(D_1, 0) = H(S) - \min_{p_{X|S} \in \calQ^{Steg}(p_S,D_1)} I(S;X) . \]
For the problem of encoding a source $S$ subject to distortion $D_1$,
the minimum rate for representing the source is given by the rate-distortion
function
\[ R_S(D_1) = \min_{p_{X|S}~:~{\mathsf E}\, \dsf(S,X) \le D_1} I(S;X)
    \le \min_{p_{X|S} \in \calQ_1^{Steg}(p_S,D_1)} I(S;X) \]
where the inequality holds because $p_{X|S} \in
\calQ_1^{Steg}(p_S,D_1)$ implies ${\mathsf E} \,\dsf(S,X) \le D_1$.
Hence
\begin{equation}
   C^{Steg}(D_1, 0) \le H(S) - R_S(D_1)
\label{eq:Csteg-RD}
\end{equation}
and the capacity-achieving codes for the passive-warden case are analogous
to rate-distortion codes.
Equality holds in (\ref{eq:Csteg-RD}) if the distribution that achieves
the rate-distortion bound satisfies the steganographic property $p_X = p_S$.
\begin{proposition}
For the passive-warden case ($D_2=0$), the random-coding exponent is given by
\begin{equation}
   E_r^{Steg}(R)=\min_{\tilde p_S \in \mathcal P_S}
\max_{p_{X|S} \in \calQ_1^{Steg}(\tilde p_S, D_1)}\left[D(\tilde
p_S||p_S)+\left|H_{\tilde p_S, p_{X|S}}(X|S)-R\right|^+\right] .
\label{eq:E-steg-passive}
\end{equation}
\label{proposition:passive error exponent}
\end{proposition}
\section{Penalty for Perfect Security}
\label{sec:loss}
The capacity expressions for public watermarking
in~\cite{Moulin_O'Sullivan_IEEE03,Somekh-Baruch_Merhav_PubWat2004}
and for steganography in (\ref{eq:C-steg}) take the same form,
except that here the maximization of $p_{XU|S}$ is subject
to the steganographic constraint. Consequently, we have
\begin{equation}
C^{Steg} \le C^{PubWM} \label{eq:C ineq}
\end{equation}
and similarly
\begin{equation}
E_r^{Steg}(R) \le E_r^{PubWM}(R) . \label{eq:E ineq}
\end{equation}
For some special cases, it is possible
that the optimal covert channel for public watermarking
automatically satisfies the perfect security condition, and
(\ref{eq:C ineq}) and (\ref{eq:E ineq}) hold with equality.
Proposition~\ref{proposition:q-ary} below states sufficient
conditions on the covertext PMF $p_S$ and the distortion function
$\dsf(\cdot,\cdot)$ that ensure the perfect security constraint
causes no penalty in communication performance.

We consider $\calS=\mathbb Z_q=\{0,\,1,\,2,\,\cdots,\,q-1\}$,
which is a group under addition modulo $q$.
We shall use the notation $\underline {k} \triangleq k \mod q$.
The covertext $S$ is uniformly distributed over $\mathbb Z_q$, i.e.,
\[p_S=\mathbb U(\calS).\]
The associated distortion function $\dsf: \calS \times \calS \to
\mathbb R^+ \cup \{0\}$ satisfies
\[\dsf(i,\,i)=0 \mbox{ and } \dsf(i,\,j)=\dsf(0,\,\underline{j-i}),\]
If we write
$\{\dsf(i,\,j)\}_{i,\,j=0}^{q-1}$ in a matrix form, the
distortion matrix is cyclic-Toeplitz.
\begin{definition}
Let $\calV \triangleq \{0,\, 1,\, \cdots, \,L-1 \}$, $p_S=\mathbb
U(\calS)$, and $\calU \triangleq \{0,\,1,\,2,\,\cdots, \,qL-1\}$.
Given any covert channel $p_{XV|S} \in \calQ(L, p_S,D_1)$, where $v \in
\calV$, we define an associated covert channel $p_{XU|S} \in
\calP_{XU|S}$, where $U \in \calU$, by
\begin{equation}
p_{XU|S}\left(x,\,qv+i\big|s\right)=\frac{1}{q}\,p_{XV|S}
\left({\underline{x-i}},\,v\big|{\underline{s-i}}\right),
\; \forall \,v \in \calV,\;\forall\,i, \;s, \;x \in \calS.
\label{eq:pxu|s steg}
\end{equation}
\label{def:pxu|s}
\end{definition}

For any stochastic matrix $p_{XV|S}\in \calQ(L, p_S,D_1)$, by
(\ref{eq:pxu|s steg}), the new channel $p_{XU|S}$ contains all of
its $q$ cyclically shifted versions (with respect to $X$ and $S$)
and these shifted versions are equally likely. Since the
distortion function is cyclic, 
it is easy to verify that
\[\mathsf E_{p_S,p_{XU|S}}[\dsf(S,X)]=\mathsf E_{p_S,p_{XV|S}}[\dsf(S,X)] \le D_1.\]
Moreover, the marginal PMF $\hat p_X$ induced by $p_S=\mathbb
U(\calS)$ and $p_{XU|S}$ is given by
\begin{equation}
\hat p_X(x)=\frac{1}{q}
\sum_{i=0}^{q-1}p_X(\underline{x-i})=\frac{1}{q} \equiv
p_S(x), \quad \forall \, x \in \calS,
\end{equation}
where $p_X$ is the marginal PMF induced by $p_S=\mathbb U(\calS)$ and
$p_{XV|S} \in \calQ(L,p_S,D_1)$. That is,
\[p_{XU|S} \in \calQ^{Steg}(qL, p_S,D_1).\]
\begin{definition}
The class $\calQ_{cyc}^{Steg}(qL, p_S,D_1)$ is the set of all such
$p_{XU|S}$ defined in (\ref{eq:pxu|s steg}). \label{def:Q1}
\end{definition}

Clearly, we have
\begin{equation}
\calQ_{cyc}^{Steg}(qL, p_S,D_1) \subset \calQ^{Steg}(qL, p_S,D_1) \subset
\calQ(qL, p_S,D_1).
\label{eq:Q nested}
\end{equation}
\begin{definition}
The class of cyclic attack channels subject to distortion $D_2$ is
defined as
\begin{eqnarray}
{\calA}_{cyc}(D_2) &\triangleq& \Big \{p_{Y|X} \in \calP_{Y|X}:\;
p_{Y|X}(y|x)=p_{Y|X}(\underline{y-x}\,|\,0\,), \quad \forall \, x,\,y \in \calS,\nonumber\\
&& \quad \quad \quad \quad \mbox{ and } \quad
\frac{1}{q}\sum_{y=0}^{q-1}p_{Y|X}(y|0)\,\dsf (y,0) \le D_2\Big
\}.\label{eq:Acyc-py|x}
\end{eqnarray}
\label{def:Acyc}
\end{definition}

Any stochastic matrix $p_{Y|X}\in {\calA}_{cyc}(D_2)$ is
cyclic-Toeplitz. Also note that for any $p_X \in \calP_X$,
\begin{equation}
{\calA}_{cyc}(D_2) \subset \calA(p_X,D_2). \label{eq:Acyc in A}
\end{equation}
\begin{proposition}
For the above $q$-ary information-hiding problem, the capacities
for both the perfectly secure steganography game and the public
watermarking game are the same. That is, the perfect security
constraint in (\ref{eq:secure}) does not cause any capacity
loss. Moreover, there is no loss of optimality in restricting the
maximization in (\ref{eq:C-L-steg}) to
$\calQ^{Steg}_{cyc}(qL, p_S,D_1)$ and the minimization to
${\calA}_{cyc}(D_2)$:
\begin{eqnarray}
C^{PubWM}(D_1,D_2) &=& C^{Steg}(D_1,D_2)\nonumber\\
&=&\lim_{L \to \infty}\max_{p_{XU|S} \in \calQ^{Steg}_{cyc}(qL,p_S,D_1)}
\min_{p_{Y|X} \in {\calA}_{cyc}(D_2)}J_L(p_S,p_{XU|S},p_{Y|X}).
\label{eq:q-ary}
\end{eqnarray}
\label{proposition:q-ary}
\end{proposition}
The proof is given in Appendix~\ref{app:proposition 2 Steg}.

\section{Example: Binary-Hamming Case}
\label{sec:binary-hamming}
We illustrate the above results through the following example,
where $\calS=\{0,1\}$, and the covertext is
Bernoulli($\frac{1}{2}$) sequence, i.e.,
\[Pr[S=1]=Pr[S=0]=\frac{1}{2}.\]
The Hamming distortion metric is used: $\dsf(x,y) = \mathds1_{\{x \ne y\} }$.
\begin{figure}[bht]
   \begin{center}
   \begin{tabular}{c}
   \includegraphics[height=7cm]{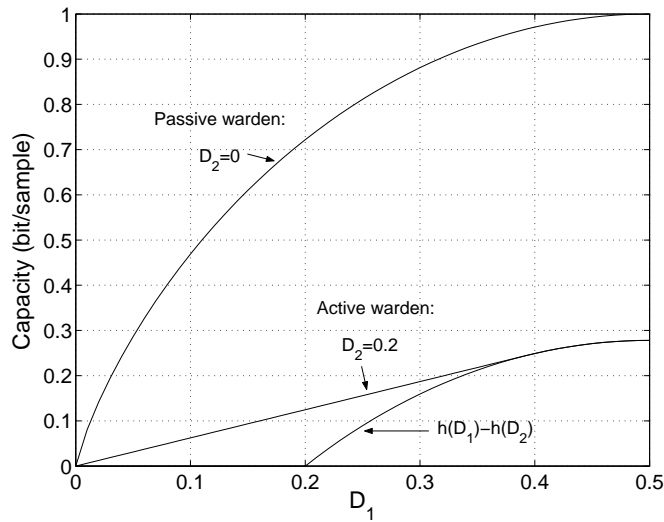}
   \end{tabular}
   \end{center}
   \caption{Capacity for a perfectly secure steganography game
    when the covertext $\bS$ is a Bernoulli($\frac{1}{2}$) sequence.}
 \label{fig:StegBinaryCapacity}
\end{figure}
\subsection{Capacity}
\label{sec:Binary-Hamming Steganographic Capacity}
The capacity in the public watermarking game setting is given
in~\cite{Moulin_Wang_IEEEIT04} as follows
\begin{equation}
C=\left \{
\begin{array}{ll}
\frac{D_1}{d_{D_2}} [h(d_{D_2})-h(D_2)],& \mbox {if} \; 0 \le D_1 \le d_{D_2} ; \\
h(D_1)-h(D_2), &\mbox {if} \; d_{D_2} \le D_1 \le 1/2;\\
1-h(D_2), &\mbox {if} \; D_1 > 1/2,
\end{array}
\right. \label{eq:binarycapacity}
\end{equation}
where $d_{D_2}=1-2^{-h(D_2)}$. 
When $D_2=0$,
\begin{equation}
C=\left \{
\begin{array}{ll}
h(D_1) & \mbox {if} \; 0 \le D_1 \le 1/2 ; \\
1 &\mbox {if} \; D_1 \ge 1/2.\\
\end{array}
\right.
\end{equation}
Fig.~\ref{fig:StegBinaryCapacity} shows the above two capacity
functions.

The optimal attack channel is a binary symmetric channel (BSC)
with crossover probability $D_2$. If $d_{D_2}
\le D_1 \le 1/2$, the optimal covert channel is also a binary
symmetric channel: BSC($D_1$) (i.e., $|\calU|=2$, $U=X$, and
$p_{XU|S}=p_{X|S}$); otherwise, the capacity is achieved by time
sharing: no embedding on a fraction of $1-\frac{D_1}{d_{D_2}}$
samples and embedding with the optimal covert channel
BSC($d_{D_2}$) on the rest of samples. Since the covertext $\bS$
is a Bernoulli($\frac{1}{2}$) sequence, the output of the above
optimal BSC($p$) covert channel is also Bernoulli($\frac{1}{2}$).
That is, the optimal covert channel for the public watermarking game
satisfies $p_X = p_S$, and the perfect security
constraint does not cause any loss in capacity, as stated by
Proposition~\ref{proposition:q-ary}.
\begin{figure}[bht]
   \begin{center}
   \begin{tabular}{c}
   \includegraphics[height=7cm]{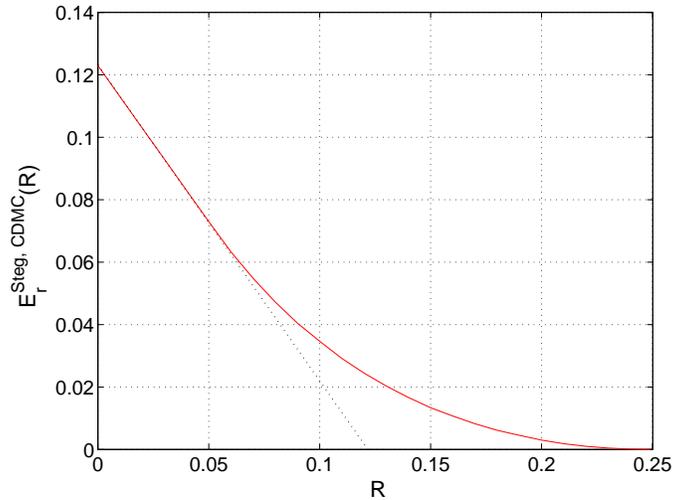}
   \end{tabular}
   \end{center}
\caption{Random-coding exponent for perfectly secure
steganography game when the covertext $\bS$ is a
Bernoulli($\frac{1}{2}$) sequence, $D_1=0.4$, $D_2=0.2$, and $|\calU|=2$.}
\label{fig:StegErrExpo}
\end{figure}
\subsection{Random-Coding Exponent}
In~\cite{Moulin_Wang_IEEEIT04}, we numerically computed the
random-coding exponent for public watermarking in the case
of $D_1=0.4$, $D_2=0.2$, and $|\mathcal U|=2$ as shown in
Fig.~\ref{fig:StegErrExpo}. We found that the optimal covert
channel is still a BSC($D_1$) ($p_{XU|S}=p_{X|S}$) with the time
sharing strategy. It implies that at least for the case of
$|\calU|=2$, $p_X = p_S$ and the perfect security constraint
causes no loss in random-coding exponent either.
\subsection{Randomized Nested Linear Codes---A Capacity-Achieving Code Construction}
\label{sec:linear codes} For information-embedding problems with a
fixed attack channel BSC($D_2$), \emph{deterministic} nested
binary linear codes were proposed to achieve capacity, where
$\calC_1$, a good source code with Hamming distance $D_1$, is
nested in $\calC_2$, a good channel code over
BSC$(D_2)$~\cite{Zamir_Shamai_Erez_IEEEIT02,Barron_Chen_Wornell_BCW03}.
When $|\calC_2|\doteq 2^{N\left[1-h(D_2)\right]}$ and $|\calC_1|
\doteq 2^{N\left[1-h(D_1)\right]}$, the asymptotic code rate
\[R=\lim_{N \to \infty}\frac{1}{N}\log_2\frac{|\calC_2|}{|\calC_1|}=h(D_1)-h(D_2)\]
is equal to the capacity in the regime $D_1 \ge d_{D_2}$. In the regime $D_1 < d_{D_2}$,
the time-sharing strategy of (\ref{eq:binarycapacity}) is applied. These nested linear
codes apply to both public watermarking and steganography because BSC($D_2$) is the
optimal discrete memoryless attack channel.
The stegotext codewords are elements of $\calC_2$
\cite{Zamir_Shamai_Erez_IEEEIT02,Barron_Chen_Wornell_BCW03}.

In the passive-warden case ($D_2=0$), we simply let $\calC_2 = \mathbb F_2^N$,
and perfect security is achieved even without a secret key.
In the active-warden case, $\calC_2$ is a subgroup of $\mathbb F_2^N$,
and randomization via the secret key plays
an essential role in achieving perfect security. The strategy described
below makes the transmitted stegotext uniformly distributed over $\mathbb F_2^N$.
The resulting code is a \emph{randomized} nested binary linear code.

Partition the whole space $\mathbb F_2^N$ into a disjoint union
of $\calC_2$ and its cosets:
\begin{equation}
\mathbb F_2^N=\bigcup_{\bc \in \Omega_2} \calC_2 \oplus \bc,
\label{eq:coset}
\end{equation}
where $\calC_2 \oplus \bc$ is a coset of $\calC_2$, the element
$\bc \in \Omega_2$ is a coset leader, and the set $\Omega_2$
contains all coset leaders. We have
\begin{equation}
|\Omega_2|=\frac{2^N}{|\calC_2|}\doteq 2^{N h(D_2)}.
\label{eq:coset size}
\end{equation}
Let the secret key $\bK$ be uniformly distributed over $\Omega_2$.
For any $\bk \in \Omega_2$, the encoder output is defined as
\begin{equation}
\bx=f_N^{\bk}(m,\bs)=f_N^{\mathbf 0}(m,\bs \oplus \bk) \oplus \bk,
\label{eq:randomized encoder}
\end{equation}
where $f_N^{\mathbf 0}(\cdot,\cdot)$ is the deterministic encoder
used for the information-embedding or watermarking problem.
The decoding function is
\begin{equation}
\hat m=\phi_N^{\bk}(\by)=\phi_N^{\mathbf 0}(\by \oplus \bk),
\end{equation}
where $\phi_N^{\mathbf 0}(\cdot)$ is the decoder associated with
$f_N^{\mathbf 0}(\cdot)$.

Since the output of the deterministic encoder is uniformly
distributed over $\calC_2$ and the secret key $\bK$ is uniformly
distributed over $\Omega_2$, the coset decomposition property (\ref{eq:coset})
ensures that the randomized encoder output
of (\ref{eq:randomized encoder}) is uniformly distributed over
$\mathbb F_2^N$. Hence perfect security is achieved.
By (\ref{eq:coset size}) the entropy rate of the
secret key is $h(D_2)$, unlike the $\log_2 N$ growth required
for general RM codes in (\ref{eq:entropy rate E RM}).

\section{Conclusion}
\label{sec:Steg Discussion}
A strict definition of perfect security has been adopted in this paper,
implying that even a warden with unlimited computational resources
is unable to reliably detect the presence of a hidden message.
We have studied the Shannon-theoretic limits of communication
performance under this perfect-security requirement and studied
the structure of codes that asymptotically achieve those limits.
The main results are summarized below.
\begin{itemize}
\item Perfectly secure steganography is closely related to the
public watermarking problem
of~\cite{Moulin_O'Sullivan_IEEE03,Moulin_Wang_IEEEIT04}. Positive
capacity and random-coding exponents are achieved using
stacked-binning codes and an MPMI decoder.
\item Randomized codes
are generally needed to achieve perfect security. The common
randomness is provided by a secret key shared between the encoder
and decoder. For i.i.d. covertexts, Proposition~\ref{prop:order1}
shows that perfectly secure steganographic codes can be
constructed using randomized permutations of a prototype CCC
watermarking code that merely has an order-1 security property,
i.e., the prototype code matches the first-order marginals of the
covertext and stegotext, but not the full $N$-dimensional statistics.
\item The cost of perfect security in terms of
communication performance is the same as the cost of order-1
security. However, if the covertext distribution is uniform and
the distortion metric is cyclically symmetric, the security
constraint does not cause any loss of performance.
\end{itemize}

{\bf Computational Security.} This paper has focused on the
interplay between communication performance and
information-theoretic security, where security is achieved using a
private key that is uniformly distributed over a group
$\calG^{sub}$. A more practical setup would involve a public-key
system, in which a reduced set of representers of $\calG^{sub}$ is
selected, each corresponding to a value of the key. Assume that
the uniform distribution over this reduced set is computationally
indistinguishable (in a sense to be precisely defined) from the
uniform distribution over $\calG^{sub}$. The resulting
steganographic code is no longer perfectly secure but inherits the
computational security of the key generation mechanism. Thus the
framework analyzed in this paper can form the basis for
constructing computationally secure steganographic codes that have
near-optimal communication performance.

{\bf Extensions.} Our basic framework can also be used to analyze
complex problems involving covertexts with Markov dependencies and
covertexts defined over continuous alphabets \cite[Sec.
X]{Wang-Moulin-arxiv07}. While such extensions are technically
challenging, we hope that the mathematical structure of optimal
codes identified in this paper under simplifying assumptions will
shed some light on the development of practical codes with high
communication performance. \clearpage \appendices

\section{Proof of Prop~\ref{prop:order1}}
\label{app:order1}
First we verify the perfect security
condition. For RM codes (Def.~\ref{def:RM codes}), we have
\[ p_{\bX|\pi,\bS,M}(\bx|\pi,\bs,m) = \mathds1_{\{ \pi \bx = f_N(\pi\bs,m)\} } . \]
Also note that for any $\bx, \bz \in T_{\bs}$, there exists a permutation $\pi_0$
such that $\bx = \pi_0 \bz$. Hence the value of the sum
$\sum_{\pi} \mathds1_{\{ \pi \bx = \bz\} }$ is independent of $\bz$
(conditioned on $\bz \in T_{\bs}$), and so
\begin{equation}
   \sum_{\pi} \mathds1_{\{ \pi \bx = \bz\} }
    = \frac{1}{|T_{\bs}|} \sum_{\bz \in T_{\bs}} \sum_{\pi} \mathds1_{\{ \pi \bx = \bz\} }
    = \frac{1}{|T_{\bs}|} \sum_{\pi} 1
    = \frac{N!}{|T_{\bs}|} .
\label{eq:sum-pi}
\end{equation}
Hence for any type class $T_{\bs}$ we have
\begin{eqnarray}
   p_{\bX|T_{\bs}}(\bx|T_{\bs})
    & = & \frac{1}{N!} \sum_{\pi} \frac{1}{|\calM|} \sum_{m \in \calM}
        \frac{1}{|T_{\bs}|} \sum_{\bs' \in T_{\bs}}
        p_{\bX|\pi,\bS,M}(\bx|\pi,\bs',m) \nonumber \\
    & = & \frac{1}{N!} \sum_{\pi} \frac{1}{|\calM|} \sum_{m \in \calM}
        \frac{1}{|T_{\bs}|} \sum_{\bs' \in T_{\bs}} \mathds1_{\{ \pi \bx = f_N(\pi\bs',m)\} }
                                        \nonumber \\
    & \stackrel{(a)}{=} & \frac{1}{N!} \sum_{\pi} \frac{1}{|\calM|} \sum_{m \in \calM}
        \frac{1}{|T_{\bs}|} \sum_{\bs'' \in T_{\bs}} \mathds1_{\{ \pi \bx = f_N(\bs'',m)\} }
                                        \nonumber \\
    & = & \frac{1}{|\calM|} \sum_{m \in \calM} \frac{1}{|T_{\bs}|} \sum_{\bs'' \in T_{\bs}}
        \frac{1}{N!} \sum_{\pi} \mathds1_{\{ \pi \bx = f_N(\bs'',m)\} } \nonumber \\
    & \stackrel{(b)}{=} & \frac{1}{|\calM|} \sum_{m \in \calM} \frac{1}{|T_{\bs}|}
        \sum_{\bs'' \in T_{\bs}} \frac{1}{|T_{\bs}|} \,\mathds1_{\{\bx \in T_{\bs}\} }
                                        \nonumber \\
    & = & \frac{1}{|T_{\bs}|} \,\mathds1_{\{\bx \in T_{\bs}\} },
\label{eq:px-Ts}
\end{eqnarray}
where in (a) we have made the change of variables $\bs'' = \pi \bs'$,
and in (b) we have used (\ref{eq:sum-pi}) with $\bz = f_N(\bs'',m)$.
From (\ref{eq:px-Ts}) we obtain
\[ p_{\bX}(\bx) = \sum_{T_{\bs}} p_{\bS}(T_{\bs}) \,p_{\bX|T_{\bs}}(\bx|T_{\bs})
    = \sum_{T_{\bs}} p_{\bS}(T_{\bs}) \,\frac{1}{|T_{\bs}|} \,\mathds1_{\{\bx \in T_{\bs}\} }
    = p_{\bS}(\bx) , \quad \forall \bx \in \calS^N ,
\]
hence the perfect security condition (\ref{eq:secure}) is satisfied.

Now verifying the maximum-distortion constraint (\ref{eq:D1}),
for every $\pi$ we have
\[ \dsf^N(\bs,f_N^{\pi}(\bs,m)) \stackrel{(a)}{=} \dsf^N(\bs,\pi^{-1} f_N(\pi\bs,m))
    \stackrel{(b)}{=} \dsf^N(\pi\bs,f_N(\pi\bs,m)) \stackrel{(c)}{\le} D_1 \]
where (a) uses the definition of $f_N^{\pi}$ in (\ref{eq:rm-f}),
(b) holds because the distortion measure is additive, and
(c) holds because of our initial assumption on the prototype $f_N$.
Therefore (\ref{eq:D1}) holds.

Finally, let us evaluate the error probability for the RM code.
Since the covertext source and the attack channel are memoryless,
we have
\begin{equation}
   p_S^N(\bs) = p_S^N(\pi\bs) \quad \mathrm{and} \quad
    p_{Y|X}^N(\by|\bx) = p_{Y|X}^N(\pi\by|\pi\bx)
\label{eq:memoryless}
\end{equation}
for any permutation $\pi$.
The error probability for the prototype code takes the form
\begin{eqnarray*}
P_{e,N}(f_N,\phi_N,p_{Y|X})
& = & \frac{1}{|\calM|} \sum_{m \in \calM} \sum_{\bs \in \calS^N} p_S^N(\bs)
    \sum_{\bx \in \calS^N} \mathds1_{\{ \bx=f_N(\bs,m)\} }
    \sum_{\by \in \calS^N} p_{Y|X}^N(\by|\bx) \,\mathds1_{\{ \phi_N(\by) \ne m\} } .
\end{eqnarray*}
For the prototype code modulated with permutation $\pi$, we have
\begin{eqnarray}
P_{e,N}(f_N^{\pi},\phi_N^{\pi},p_{Y|X})
& = & \frac{1}{|\calM|} \sum_{m \in \calM} \sum_{\bs \in \calS^N} p_S^N(\bs)
    \sum_{\bx \in \calS^N} \mathds1_{\{ \pi\bx=f_N(\pi\bs,m)\} }
    \sum_{\by \in \calS^N} p_{Y|X}^N(\by|\bx)
    \,\mathds1_{\{ \phi_N(\pi\by) \ne m\} } \nonumber \\
& \stackrel{(a)}{=} & \frac{1}{|\calM|} \sum_{m \in \calM} \sum_{\bs \in \calS^N}
    p_S^N(\pi\bs) \sum_{\bx \in \calS^N} \mathds1_{\{ \pi\bx=f_N(\pi\bs,m)\} }
    \sum_{\by \in \calS^N} p_{Y|X}^N(\pi\by|\pi\bx)
    \,\mathds1_{\{ \phi_N(\pi\by) \ne m\} } \nonumber \\
& \stackrel{(b)}{=} & \frac{1}{|\calM|} \sum_{m \in \calM} \sum_{\pi^{-1}\bs' \in \calS^N}
    p_S^N(\bs') \sum_{\pi^{-1}\bx' \in \calS^N} \mathds1_{\{ \bx'=f_N(\bs',m)\} }
    \sum_{\pi^{-1}\by' \in \calS^N} p_{Y|X}^N(\by'|\bx')
    \,\mathds1_{\{ \phi_N(\by') \ne m\} } \nonumber \\
& \stackrel{(c)}{=} & \frac{1}{|\calM|} \sum_{m \in \calM}
    \sum_{\bs' \in \calS^N} p_S^N(\bs')
    \sum_{\bx' \in \calS^N} \mathds1_{\{ \bx'=f_N(\bs',m)\} }
    \sum_{\by' \in \calS^N} p_{Y|X}^N(\by'|\bx')
    \,\mathds1_{\{ \phi_N(\by') \ne m\} } \nonumber \\
& = & P_{e,N}(f_N,\phi_N,p_{Y|X}),
\label{eq:fNpi-prob}
\end{eqnarray}
where (a) holds because of (\ref{eq:memoryless}),
(b) is obtained using the change in variables $\bs'=\pi\bs$, $\bx'=\pi\bx$, $\by'=\pi\by$,
and (c) holds because the three sums run over all elements ($\bs',\bx',\by'$) of
$\calS^N \times \calS^N \times \calS^N$, and so the order of summation
is inconsequential.
Since (\ref{eq:fNpi-prob}) holds for every permutation $\pi$,
the error probability for the RM code
is equal to
\[ P_{e,N}(F_N,\Phi_N,p_{Y|X})
    = \frac{1}{N!} \sum_{\pi} P_{e,N}(f_N^{\pi},\phi_N^{\pi},p_{Y|X})
    = P_{e,N}(f_N,\phi_N,p_{Y|X}) .
\]
This completes the proof.

\section{Converse Proof of Theorem~\ref{theorem:Stegcapacity}}
\label{app:Stegconverse}
The converse is an extension of the proof in \cite[Section
7]{Moulin_Wang_IEEEIT04}. Our upper bound on achievable rates
is derived by
\begin{itemize}
\item replacing the perfect-security constraint with a weaker order-1
    security constraint on the encoder:
    \begin{equation}
       p_{\bx} = p_{\bs} \quad \forall \, m, \bs, \bx=f_N(\bs,m)
    \label{eq:order1-converse}
    \end{equation}
    (matching the types of input $\bs$ and output $\bx=f_N(s,m)$ of the encoder
    $f_N$),
\item replacing the almost-sure distortion constraint with an expected
    distortion constraint on the encoder:
    \begin{equation}
       \frac{1}{|\calM|} \sum_{\bs \in \calS^N} p_S^N(\bs)
       \,\dsf^N(\bs,f_N(\bs,m)),
    \label{eq:D1-expected}
    \end{equation}
\item and providing the decoder with knowledge of the attack channel $p_{Y|X}$.
\end{itemize}
Clearly any upper bound we derive under these assumptions
is an upper bound on capacity as well.

For any rate-$R$ code $(f_N,\phi_N)$ and DMC
$p_{Y|X} \in \calA(p_X,D_2)$, we have
\begin{eqnarray}
NR=H(M)&=&H(M|\bY)+I(M;\bY) \nonumber\\
&\le&1+P_e(f_N,\phi_N,p^N_{Y|X})\,NR+I(M;\bY),\nonumber
\end{eqnarray}
where the inequality is due to Fano's inequality. In order for
$P_e$ not to be bounded away from 0, rate $R$ needs to satisfy
\begin{eqnarray}
NR-1 \le \min_{p_{Y|X} \in \calA(p_X,D_2)} I(M;\bY).
\label{eq:nr<Imy}
\end{eqnarray}

The joint PMF of $(M,\bS, \bX, \bY)$ is given by
\begin{eqnarray}
p_{M\bS\bX\bY|f_N}=p_M \,p_S^N\, p_{Y|X}^N
\,\mathds1_{\{\bX=f_N(\bS,M)\} }. \label{eq:pmsxyfN}
\end{eqnarray}

Owing to (\ref{eq:pmsxyfN}), for any $1\le i \le N$,
$(M,\bS,\{Y_j\}_{j \neq i})\to X_i \to Y_i$ forms a Markov chain
and so does
\begin{eqnarray}
(W_i, S_i) \to X_i\to Y_i,
\end{eqnarray}
where the random variable $W_i$ is defined as
\begin{eqnarray}
W_i=(M,S_{i+1},\cdots,S_N,Y_1, \cdots, Y_{i-1}).
\label{eq:Wi}
\end{eqnarray}
Using the same set of inequalities as in~\cite[Lemma
4]{Gel'fand_Pinsker_1980}, we obtain
\begin{eqnarray}
I(M;\bY)\le \sum_{i=1}^N \,[I(W_i;Y_i)-I(W_i;S_i)].
\label{eq:Imy<sumIwy-Iws}
\end{eqnarray}

We define a time sharing random variable $T$, which is uniformly
distributed over $\{1, \cdots, N\}$ and independent of all other
random variables, and define the quadruple of random variables
$(W,S,X,Y)$ as $(W_T,S_T, X_T,Y_T)$. With this definition, the
order-1 security constraint (\ref{eq:order1-converse}) becomes
$p_X = p_S$, and the expected distortion constraint
(\ref{eq:D1-expected}) becomes $\sum_{s,x} p_S(s) p_{X|S}(x|s)
\,\dsf(s,x) \le D_1$. Therefore $p_{X|S} \in
\calQ_1^{Steg}(p_S,D_1)$.

By (\ref{eq:Wi}), the random variable $W$ is defined over an
alphabet of cardinality $\exp_2\left\{N\left[R+\log |S|
\,\right]\right\}$. Moreover $(W, S) \to X \to Y$ forms a Markov
chain. Combining (\ref{eq:nr<Imy}) and (\ref{eq:Imy<sumIwy-Iws}),
we further derive
\begin{eqnarray}
R &\le& \frac{1}{N}\min_{p_{Y|X} \in \calA(p_X,D_2)}I(M;\bY)\nonumber\\
&\le& \frac{1}{N} \min_{p_{Y|X} \in \calA(p_X,D_2)}
    \sum_{i=1}^N [I(W_i;Y_i)-I(W_i;S_i)]\nonumber\\
&=&\min_{p_{Y|X} \in \calA(p_X,D_2)}\left[I(W;Y|T)-I(W;S|T)\right]\nonumber\\
&=&\min_{p_{Y|X} \in \calA(p_X,D_2)}\left[I(W,T;Y)-I(W,T;S)-I(T;Y)+I(T;S)
    \right]\nonumber\\
&\le&\min_{p_{Y|X} \in \calA(p_X,D_2)}\left[I(U;Y)-I(U;S)\right],
\label{eq:R converse}
\end{eqnarray}
where $U=(W,T)$ is defined over an alphabet of cardinality
\begin{equation}
L(N)=N\exp_2\{N\left[R+\log |S| \,\right]\},
\end{equation}
and the last inequality is due to $I(T;Y)\ge0$ and $I(T;S)=0$
(since $T$ is independent of $S$). Since $p_{X|S} \in
\calQ_1^{Steg}(p_S,D_1)$, we have $p_{XU|S} \in
\calQ^{Steg}(L(N),p_S,D_1)$.

Recall that $J_L(p_S,p_{XU|S}, p_{Y|X})\triangleq I(U;Y)-I(U;S)$
when $|\calU|=L$, and that
\[ C_L^{Steg} \triangleq \max_{p_{XU|S} \in \calQ^{Steg}(L,p_S,D_1)}
    \min_{p_{Y|X} \in \calA(p_X,D_2)}J_L(p_S,p_{XU|S}, p_{Y|X}).\]
Following the same arguments as in \cite{Moulin_Wang_IEEEIT04},
the sequence $C_L^{Steg}$ is nondecreasing and converges
to a finite limit
\[C^{Steg}\triangleq \lim_{L\to\infty} C_L^{Steg}=\lim_{L\to\infty}
    \max_{p_{XU|S} \in \calQ^{Steg}(L,p_S,D_1)}\min_{p_{Y|X} \in \calA(p_X,D_2)}
    J_L(p_S,p_{XU|S}, p_{Y|X}).\]

Therefore, continuing with (\ref{eq:R converse}), $R$ is bounded
by
\begin{eqnarray}
R&\le&\min_{p_{Y|X} \in \calA(p_X,D_2)}\left[I(U;Y)-I(U;S)\right]\nonumber\\
&=&\min_{p_{Y|X} \in \calA(p_X,D_2)}J_{L(N)}(p_X,p_{XU|S}, p_{Y|X})\nonumber\\
&\le&\sup_L \max_{p_{UX|S} \in \calQ^{Steg}(L,p_S,D_1)} \min_{p_{Y|X}
\in
\calA(p_X,D_2)}J_L(p_S,p_{XU|S}, p_{Y|X})\nonumber\\
&=&\lim_{L\to\infty}\max_{p_{XU|S} \in
\calQ^{Steg}(L,p_S,D_1)}\min_{p_{Y|X} \in \calA(p_X,D_2)}
J_L(p_S,p_{XU|S}, p_{Y|X})\nonumber\\
&=&C^{Steg}. \label{eq:Imy<N(Iuy-Ius)}
\end{eqnarray}
This proves the converse part of Theorem~\ref{theorem:Stegcapacity}.

\section{Proof of Proposition~\ref{proposition:exponent}}
\label{app:Stegdirect}
We have
\[ E_{r,L}^{Steg}(R) \le E_{r,L}^{PubWM}(R) . \]
Recall from~\cite[Lemma~3.1]{Moulin_Wang_IEEEIT04} that the
sequence $E_{r,L}^{PubWM}(R)$ is nondecreasing and converges to a
finite limit $E_{r}^{PubWM}(R)$ as $L \to \infty$. Using the same
arguments as in \cite[Lemma~3.1]{Moulin_Wang_IEEEIT04}, it follows
that the sequence $E_{r,L}^{Steg}(R)$ is nondecreasing and
converges to a finite limit $E_{r}^{Steg}(R)$ as $L \to \infty$.
Hence for any $\epsilon >0$ and $R$, there exists $L(\epsilon)$
such that
\[E_{r,L}^{Steg}(R) \ge E_r^{Steg}(R)-\epsilon, \quad \forall \,L \ge L(\epsilon).\]

We next prove that for any $L$, a sequence of {\em deterministic}
codes $(f_N,\phi_N)$ with order-1 steganographic security exist
with the property that
\[\lim_{N \to \infty}\left[-\frac{1}{N}\log \max_{p_{Y|X} \in \calA(p_X,D_2)}
    P_e(f_N,\phi_N,p_{Y|X})\right] = E_{r,L}^{Steg}(R).\]
To prove the existence of such a code, we construct a random ensemble $\mathscr C$
of binning codes $(f_N,\phi_N)$ with auxiliary alphabet
$\mathcal U \triangleq\{1,2,\cdots,L\}$ and show that the error probability
averaged over $\mathscr C$ vanishes at rate $E_{r,L}^{Steg}(R)$ as $N$ goes
to infinity. The proof is based on that of \cite[Theorem~3.2]{Moulin_Wang_IEEEIT04}
with special treatment on the encoder construction for perfect
security.

Assume that $R < C_L^{Steg} -\epsilon$. For any covertext type
$p_\bs$ and conditional type $p_{\bx \bu|\bs}$, define the
function
\begin{eqnarray}
E_{L,N}(R,p_\bs,p_{\bx\bu|\bs}) & \triangleq &
\min_{p_{\by|\bx\bu\bs}} \min_{p_{Y|X} \in
\calA(p_X,D_2)}\Big[D(p_\bs \,p_{\bx\bu|\bs}\,
p_{\by|\bx\bu\bs}||p_S \,p_{\bx\bu|\bs}\,p_{Y|X})\nonumber\\
&&\quad\quad\quad\quad\quad\quad\quad\quad\quad\quad
\quad+|I(\bu;\by)-I(\bu;\bs)-\epsilon-R|^+\Big].
\label{eq:optimal encoder}
\end{eqnarray}
Define $\calQ^{Steg}(N,L,p_{\bs},D_1)$ as the set of conditional types
$p_{\bx|\bu\bs}$ that also belong to the set $\calQ^{Steg}(L,p_{\bs},D_1)$
of feasible steganographic channels.
If $p_{\bx|\bu\bs} \in \calQ^{Steg}(N,L,p_{\bs},D_1)$ then
\begin{itemize}
    \item[(1)] $p_\bx = p_\bs$, i.e., the stegotext sequence
    has the same type as the covertext sequence and the order-1
    security condition is satisfied;
    \item[(2)] $\dsf^N(\bx,\bs) \le D_1$,
    i.e., distortion is no greater than $D_1$ for any choice of $\bs$ and $m$.
\end{itemize}
The set $\calQ^{Steg}(N,L,p_{\bs},D_1)$ includes
$p_{\bx|\bu\bs}=\mathds1_{\{\bx=\bs\}}$ and is therefore nonempty.

Now denote by $p_{\bx|\bu\bs}$ the maximizer of (\ref{eq:optimal
encoder}) over the set $\calQ^{Steg}(N,L,p_{\bs},D_1)$. As a
result of this optimization, we may associate
\begin{itemize}
\item to any covertext type $p_\bs$, a type class $T^*_U(p_\bs) \triangleq T_\bu$
and a mutual information $I^*_{US}(p_\bs) \triangleq I(\bu;\bs)$;
\item to any covertext sequence $\bs$, a conditional type class
$T^*_{U|S}(\bs) \triangleq T_{\bu|\bs}$;
\item to any sequences $\bs$ and $\bu \in T^*_{US}(p_\bs)$,
a conditional type class $T^*_{X|US}(\bu,\bs) \triangleq T_{\bx|\bu\bs}$.
\end{itemize}

A random codebook $\calC$ is the union of codeword arrays
$\calC(p_\bs)$ indexed by the covertext sequence type $p_\bs$. Let
$\rho(p_\bs) \triangleq I^*_{US}(p_\bs)+\epsilon$. The codeword
array $\calC(p_\bs)$ is obtained by drawing $2^{N(R+\rho(p_\bs))}$
random vectors independently and uniformly from the corresponding
type class $T^*_U(p_\bs)$, and arranging them in an array with
$2^{N \rho(p_\bs)}$ rows and $2^{NR}$ columns indexed by messages.
\subsection{Encoder $f_N$}
\label{sec:encoder}
Given a codebook $\calC$, a covertext sequence $\bs$, and a
message $m$, the encoder finds in $\calC(p_\bs)$ an $l$ such that
$\bu(l,m) \in T^*_{U|S}(\bs)$. If more than one such $l$ exists,
pick one of them randomly (with uniform distribution). Let
$\bu=\bu(l,m)$. If no such $l$ is available, the encoder declares an error
and draws $\bu$ from the uniform distribution over the conditional type
class $T^*_{U|S}(\bs)$. Then $\bx$ is drawn from the uniform distribution
over the conditional type class $T^*_{X|US}(\bu,\bs)$.
Recalling the discussion below (\ref{eq:optimal encoder}), $f_N$ satisfies
both the order-1 steganographic security constraint and the maximum distortion
constraint.
\subsection{Decoder $\phi_N$}
\label{sec:decoder}
Given $\by$ and the same codebook $\calC$ used by the encoder, the
decoder first seeks a covertext type $p_{\bs}$ and $\hat \bu \in
\calC(p_\bs)$ that maximizes the \emph{penalized mutual
information} criterion
\begin{equation}
\!\!\max_{p_\bs} \max_{\bu \in\calC(p_\bs)}
[I(\bu;\by)-\rho(p_\bs)]. \label{eq:DecodingRule-PMI}
\end{equation}
The decoder then outputs the column index $\hat m$ that
corresponds to $\hat \bu$. If there exist maximizers with more
than one column index, the decoder declares an error.
\subsection{Error Probability Analysis}
\label{sec:error probability analysis}
The probability of error is given by
\[P_{e,N} \triangleq \max_{p_{Y|X} \in \calA(p_X,D_2)} Pr(M \neq \hat M)
    =\max_{p_{Y|X} \in \calA(p_X,D_2)} P_e(f_N,\phi_N,p_{Y|X}).\]
Following the steps in~\cite[Section 5]{Moulin_Wang_IEEEIT04}, the encoding
error vanishes double-exponentially and only the decoding error contributes
to $P_{e,N}$ on the exponential scale:
\begin{eqnarray}
P_{e,N} \dotle \exp_2\left\{-N \min_{p_\bs}
\max_{p_{\bx\bu|\bs}}E_{L,N}(R,p_\bs,p_{\bx\bu|\bs})\right\}.
\end{eqnarray}
As $N \to \infty$, by~\cite[Lemma~2.2]{Moulin_Wang_IEEEIT04}, the
above error exponent converges to
\begin{eqnarray}
\!\!\!\!\!\!\!\!\!\!\!\!\!\!\!\!\!\!E_{r,L}^{Steg}(R)
&\!\!\!\!=\!\!\!\!&\min_{\tilde p_S \in \calP_S}
\max_{p_{XU|S} \in \calQ^{Steg}(L,\tilde p_S, D_1)}\min_{\tilde p_{Y|XUS} \in
\calP_{Y|XUS}}\min_{p_{Y|X} \in
\calA(p_X,D_2)}\nonumber\\
&&\!\!\!\!\left[D(\tilde p_S \,p_{XU|S}\, \tilde p_{Y|XUS}||p_S
\,p_{XU|S}\, p_{Y|X})+\left|J_L(\tilde p_S, p_{XU|S}, \tilde
p_{Y|XUS})-R\right|^+\right]. \label{eq:Er-1}
\end{eqnarray}

Clearly, $E_{r,L}^{Steg}(R) \ge 0$, with equality if and
only if the following conditions are met:
\begin{itemize}
    \item the minimizing PMF $\tilde p_S$ is equal to $p_S$;
    \item the minimizing conditional PMF $\tilde p_{Y|XUS}$ is equal to $p_{Y|X}$; and
    \item $R \ge \max_{p_{XU|S} \in \calQ^{Steg}(L,p_S,D_1)}
        \min_{p_{Y|X} \in \calA(p_X,D_2)}J_L(p_S,p_{XU|S},p_{Y|X}) =C_L^{Steg}$.
\end{itemize}
Therefore, $E_{r,L}^{Steg}(R) > 0$ and the error probability
vanishes for any $R<C_L^{Steg}(D_1,D_2)$. This implies that the capacity is
lower-bounded by
\[ \lim_{L \to \infty}C_L^{Steg}(D_1,D_2). \]
\subsection{Perfect Security}
Having established the achievability of $E_{r,L}^{Steg}(R)$ and
$C_L^{Steg}$ for a deterministic code $(f_N,\phi_N)$ with order-1
security and maximum distortion $D_1$, we invoke
Proposition~\ref{prop:order1} to claim that the randomly modulated
code with prototype $(f_N,\phi_N)$ achieves the same error
probability (hence error exponent) and distortion as the
prototype.
\section{Proof of Proposition~\ref{proposition:passive capacity}}
\label{app:C-passive}
By (\ref{eq:passive}), $J_L(p_S,p_{XU|S},p_{Y|X})$ is reduced to
\[J_L(p_S,p_{XU|S},p_{Y|X})=I(U;X)-I(U;S).\]
Coosing $U=X$ yields the lower bound
\begin{eqnarray}
C^{Steg}(D_1, 0) &\ge& \max_{p_{X|S} \in \calQ_1^{Steg}(p_S,D_1)}
I(X;X)-I(X;S) \nonumber\\
&=&\max_{p_{X|S} \in \calQ_1^{Steg}(p_S,D_1)} H(X|S).
\label{eq:private proof-lb}
\end{eqnarray}
On the other hand,
\begin{eqnarray}
J_L(p_S,p_{XU|S},p_{Y|X})
&=&I(U;X)-I(U;S) \nonumber\\
&\le&I(U;X|S)\label{eq:private proof-1}\\
&=& H(X|S)-H(X|U,S) \nonumber \\
&\le&H(X|S).\label{eq:private proof}
\end{eqnarray}
Note that (\ref{eq:private proof-1}) follows from the chain rule
of mutual information
\[I(U;XS)=I(U;X)+I(U;S|X)=I(U;S)+I(U;X|S)\]
and $I(U;S|X)\ge 0$. Choosing $U=X$ achieves equality in both
(\ref{eq:private proof-1}) and (\ref{eq:private proof}).

From (\ref{eq:private proof}), we obtain
\begin{eqnarray}
C^{Steg}(D_1, 0) &=& \lim_{L \to \infty}\max_{p_{XU|S} \in
\calQ^{Steg}(L,p_S,D_1)}J_L(p_S,p_{XU|S},p_{Y|X})\nonumber\\
&\le& \lim_{L \to \infty}\max_{p_{XU|S} \in
\calQ^{Steg}(L,p_S,D_1)}H(X|S)\nonumber\\
&=&\max_{p_{X|S} \in \calQ^{Steg}(p_S,D_1)}H(X|S). \label{eq:private
proof-ub}
\end{eqnarray}

Combining (\ref{eq:private proof-lb}) and (\ref{eq:private
proof-ub}) yields (\ref{eq:C-steg-passive}) and proves the
proposition.
\section{Proof of Proposition~\ref{proposition:passive error exponent}}
\label{app:E-passive}
Since $p_{Y|X}=\mathds1_{\{Y=X\} }$, the term
$D(\tilde p_S \,p_{XU|S}\, \tilde p_{Y|XUS}||p_S\, p_{XU|S}\, p_{Y|X})$
in (\ref{eq:ErL}) is
infinite if $\tilde p_{Y|XUS}\not = p_{Y|X}$. Hence, the
minimizing $\tilde p_{Y|XUS}$ in (\ref{eq:ErL}) is given by
\[\tilde p^*_{Y|XUS}=p_{Y|X}= \mathds1_{\{Y=X\} }.\]
Consequently, the two terms of the cost function of (\ref{eq:ErL})
are reduced to
\[D(\tilde p_S\, p_{XU|S} \,\tilde
p^*_{Y|XUS}||p_S\, p_{XU|S}\, p_{Y|X})=D(\tilde p_S||p_S)\] and
\[\left|J_L(\tilde p_S, p_{XU|S}, \tilde
p^*_{Y|XUS})-R \right|^+=\left|J_L(\tilde p_S, p_{XU|S},
p_{Y|X})-R \right|^+,\] respectively. This yields
\begin{eqnarray}
\!\!\!\!\!\!\!\!\!\!\!\!\!\!\!E_{r}^{Steg}(R)
\!\!\!&=&\!\!\!\!\!\min_{\tilde p_S \in \mathcal P_S}
\left[D(\tilde p_S||p_S)+\lim_{L \to \infty}\max_{p_{XU|S} \in
\calQ^{Steg}(L,\tilde p_S, D_1)}|J_L(\tilde p_S, p_{XU|S},
p_{Y|X})-R|^+\right]. \;\; \;\label{eq:passive expo 1}
\end{eqnarray}
Similarly to the steps in the proof of
Proposition~\ref{proposition:passive capacity}, we derive that
\begin{equation}
\forall \; L \ge 2: \; \max_{p_{XU|S} \in \calQ^{Steg}(L,\tilde p_S, D_1)}
    |J_L(\tilde p_S, p_{XU|S},p_{Y|X})-R|^+
    =\max_{p_{X|S} \in \calQ_1^{Steg}(p_S,D_1)}|H_{\tilde p_S,p_{X|S}}(X|S)-R|^+.
\label{eq:passive expo 2}
\end{equation}
The maximum on the left side is achieved by $U=X$. Combining
(\ref{eq:passive expo 1}) and (\ref{eq:passive expo 2}) proves the
proposition.
\section{Proof of Proposition~\ref{proposition:q-ary}}
\label{app:proposition 2 Steg}
We prove Proposition~\ref{proposition:q-ary} in two parts. We
first establish that the right-hand side of (\ref{eq:q-ary}) is an
upper bound on the public watermarking capacity $C^{PubWM}$. Then
we prove that the right-hand side of (\ref{eq:q-ary}) is at the
same time a lower bound on the perfectly secure steganographic
capacity $C^{Steg}$.

We start with the following lemma on the properties of $p_{XU|S}
\in \calQ^{Steg}_{cyc}(qL, p_S,D_1)$, which are used throughout this
proof.
\begin{lemma}
Any $p_{XU|S} \in \calQ^{Steg}_{cyc}(qL, p_S,D_1)$ generated by
(\ref{eq:pxu|s steg}) from its corresponding $p_{XV|S} \in \calQ
(L, p_S,D_1)$ has the following properties:\\
\emph{(i) }{ }
$p_{S|U}\big(s\big|qv+i\big)=p_{S|V}\big({\underline{s-i}}\big|v\big)$,
$\forall\,i,\,s \in \calS$ and $ \forall \, v \in \calV$;\\
\emph{(ii) }
$p_{X|U}\big(x\big|qv+i\big)=p_{X|V}\big({\underline{x-i}}\big|v\big)$,
$\forall\,i,\,x \in \calS$ and $\forall\,v \in \calV$;\\
\emph{(iii) } $p_U(qv+i)=\frac{1}{q}p_V(v)$, $\forall\,i \in
\calS$, $v\in \calV$, where $p_U$ (resp. $p_V$) is the marginal
PMF of $U$ (resp. $V$) induced from $p_{XU|S}$ (resp. $p_{XV|S}$)
and
$p_S=\mathbb U(\calS)$; and\\
\emph{(iv) } $\hat p_X=\mathbb U(\calS)$, where $\hat p_X$ is the
marginal PMF of $X$ induced from $p_{XU|S}$ and $p_S=\mathbb
U(\calS)$.
\label{lemma:pxu|s property}
\end{lemma}
It is straightforward to verify Lemma~\ref{lemma:pxu|s
property}(i)-(iv) from (\ref{eq:pxu|s steg}).
\subsection{Upper Bound}
\label{sec:upper bound}
For the capacity of the public watermarking game,
\begin{eqnarray}
C^{PubWM}(D_1,D_2)&=&\lim_{L\to\infty}\max_{p_{XV|S} \in \calQ(L, p_S,D_1)}
\min_{p_{Y|X} \in \calA(p_X,D_2)}J_L(p_S, p_{XV|S},p_{Y|X}) \nonumber\\
&\le& \lim_{L\to\infty}\max_{p_{XV|S} \in \calQ(L, p_S,D_1)}
\min_{p_{Y|X} \in \calA_{cyc}(D_2)}J_{L}(p_S,
p_{XV|S},p_{Y|X}),\label{eq:C ub-1}
\end{eqnarray}
since $\calA_{cyc}(D_2) \subset \calA(p_X,D_2)$ by (\ref{eq:Acyc
in A}).

Given any $p_{XV|S} \in \calQ(L, p_S,D_1)$ and its associated $p_{XU|S}
\in \calQ_{cyc}^{Steg}(qL, p_S,D_1)$, we first verify that
\begin{equation}
I(S;U)=I(S;V). \label{eq:Isu}
\end{equation}
From $p_S=\mathbb U(\calS)$ and $p_{XV|S}$, we obtain
\begin{eqnarray}
H(S|V)&=&-\sum_{v=0}^{L-1}p_V(v)\sum_{s=0}^{q-1}p_{S|V}(s|v)\log
p_{S|V}(s|v). \label{eq:hs|v}
\end{eqnarray}
From $p_S=\mathbb U(\calS)$ and $p_{XU|S}$, we have
\begin{eqnarray}
H(S|U)&=&-\sum_{v=0}^{L-1}\sum_{i=0}^{q-1}\sum_{s=0}^{q-1}p_U(qv+i)\,p_{S|U}(s|qv+i)\,\log
p_{S|U}(s|qv+i)\label{eq:hs|u-1}\nonumber\\
&=&-\sum_{v=0}^{L-1}\sum_{i=0}^{q-1}\sum_{s=0}^{q-1}\frac{1}{q}\,p_V(v)\,p_{S|V}\big(\underline{s-i}\big|v)\,\log
p_{S|V}\big(\underline{s-i}\big|v)\label{eq:hs|u-2}\\
&=&\frac{1}{q}\sum_{i=0}^{q-1}H(S|V)=H(S|V),
\label{eq:hs|u}
\end{eqnarray}
where (\ref{eq:hs|u-2}) is obtained by using
Lemma~\ref{lemma:pxu|s property}(i) and (iii). 
Since $I(S;U)=H(S)-H(S|U)$ and $I(S;V)=H(S)-H(S|V)$,
(\ref{eq:Isu}) follows from (\ref{eq:hs|u}).

For the pair $\left(p_{XV|S},p_{Y|X}\right) \in \calQ(L, p_S,D_1)
\times \calA_{cyc}(D_2)$ and its associated pair
$\left(p_{XU|S},p_{Y|X}\right) \in \calQ_{cyc}^{Steg}(qL,
p_S,D_1)\times \calA_{cyc}(D_2)$, we have the following lemma that
is proved in Appendix~\ref{app:proof of lemma Iyv<Iyu}.
\begin{lemma}
\begin{equation}
I_{p_S, p_{XV|S},p_{Y|X}}(Y;V)\le I_{p_S,
p_{XU|S},p_{Y|X}}(Y;U).\label{eq:Iyv<Iyu}
\end{equation}
\label{lemma:Iyv<Iyu}
\end{lemma}

From (\ref{eq:Isu}), Lemma~\ref{lemma:Iyv<Iyu}, and the definition
of $J_L$ in (\ref{eq:J-L}), we obtain
\begin{equation}
J_L(p_S, p_{XV|S},p_{Y|X}) \le J_{qL}(p_S, p_{XU|S},p_{Y|X}),
\end{equation}
which yields
\begin{eqnarray}
&&\lim_{L\to\infty}\max_{p_{XV|S} \in \calQ(L, p_S,D_1)} \min_{p_{Y|X}
\in \calA_{cyc}(D_2)}J_L(p_S, p_{XV|S},p_{Y|X})\nonumber\\
&\le& \lim_{L\to\infty}\max_{p_{XU|S} \in
\calQ_{cyc}^{Steg}(qL, p_S,D_1)} \min_{p_{Y|X} \in
\calA_{cyc}(D_2)}J_{qL}(p_S, p_{XU|S},p_{Y|X}). \label{eq:C ub-2}
\end{eqnarray}

Therefore, (\ref{eq:C ub-1}) and (\ref{eq:C ub-2}) yield
\begin{eqnarray}
C^{Steg}(D_1,D_2)&\le& C^{PubWM}(D_1,D_2)\nonumber\\
&\le&\lim_{L\to\infty}\max_{p_{XU|S} \in
\calQ_{cyc}^{Steg}(qL, p_S,D_1)} \min_{p_{Y|X} \in
{\calA}_{cyc}(D_2)}J_L(p_S, p_{XU|S},p_{Y|X}).\label{eq:C upper
bound}
\end{eqnarray}
\subsection{Lower Bound}
\label{sec:lower bound}
Using the same argument at the end of
Appendix~\ref{app:Stegconverse} for the sequence
$\{C_L^{Steg}(D_1,D_2)\}$, we can argue that the sequence
$\{C_L^{PubWM}(D_1,D_2)\}$ is also nondecreasing and bounded by
$\log |\calS|$. Therefore, $\{C_L^{PubWM}(D_1,D_2)\}$ and any of
its subsequences converge to the same limit. That is
\begin{eqnarray}
C^{PubWM}(D_1,D_2)&=&\lim_{L \to \infty}\max_{p_{XU|S} \in
\calQ(L, p_S,D_1)} \min_{p_{Y|X} \in \calA(p_X,D_2)}J_L(p_S,
p_{XU|S},p_{Y|X})\nonumber\\
&=&\lim_{L \to \infty}\max_{p_{XU|S} \in \calQ(qL, p_S,D_1)}
\min_{p_{Y|X} \in \calA(p_X,D_2)}J_L(p_S, p_{XU|S},p_{Y|X}).
\end{eqnarray}
Similarly,
\begin{eqnarray}
C^{Steg}(D_1,D_2)&=&\lim_{L \to \infty}\max_{p_{XU|S} \in
\calQ^{Steg}(L,p_S,D_1)} \min_{p_{Y|X} \in \calA(p_X,D_2)}J_L(p_S,
p_{XU|S},p_{Y|X})\nonumber\\
&=&\lim_{L \to \infty}\max_{p_{XU|S} \in \calQ^{Steg}(qL, p_S,D_1)}
\min_{p_{Y|X} \in \calA(p_X,D_2)}J_L(p_S, p_{XU|S},p_{Y|X}).
\end{eqnarray}

From (\ref{eq:Q nested}),
\[\calQ_{cyc}^{Steg}(qL, p_S,D_1) \subset
\calQ^{Steg}(qL, p_S,D_1) \subset \calQ(qL, p_S,D_1).\] Thus, we have
\begin{eqnarray}
C^{PubWM}(D_1,D_2)&\ge& C^{Steg}(D_1,D_2)\nonumber\\
& \ge& \lim_{L \to \infty}\max_{p_{XU|S} \in
\calQ_{cyc}^{Steg}(qL, p_S,D_1)} \min_{p_{Y|X} \in
\calA(p_X,D_2)}J_L(p_S, p_{XU|S},p_{Y|X}).
\end{eqnarray}

Given $p_{Y|X} \in \calA(p_X,D_2)$, we define $q$ conditional
PMFs:
\begin{eqnarray}
p^m_{Y|X}(y|x)=p_{Y|X}(\underline{y-m}|\underline{x-m})
\label{eq:py|x-m}, \quad \forall\, x,\,y \in \calS, \, 0 \le m <q.
\end{eqnarray}
Since the distortion matrix $\{\dsf(i,\,j)\}_{i,\,j=0}^{q-1}$ is
cyclic, it is easy to verify that all the $q$ conditional PMFs
$p^m_{Y|X} \in \calA (p_X,D_2)$.

The conditional PMF $p_{Y|U}^m$ induced by $\left(p_{XU|S},
p^m_{Y|X}\right)\in   \calQ_{cyc}^{Steg}(qL, p_S,D_1)\times \calA
(p_X,D_2) $ is given by
\begin{eqnarray}
p_{Y|U}^m(y|qv+i)&=&\sum_{x=0}^{q-1}p_{X|U}(x|qv+i)p_{Y|X}^m(y|x)\nonumber\\
&=&\sum_{x=0}^{q-1}p_{X|U}(x|qv+i)p_{Y|X}(\underline{y-m}|\underline{x-m})\label{eq:py|u-m-1}\\
&=&\sum_{x=0}^{q-1}p_{X|V}(\underline{x-i}|v)p_{Y|X}(\underline{y-m}|\underline{x-m})\label{eq:py|u-m-2}\\
&=&\sum_{x=0}^{q-1}p_{X|U}(\underline{x-m}|qv+\underline{i-m})p_{Y|X}(\underline{y-m}|\underline{x-m})\label{eq:py|u-m-3}\\
&=&p_{Y|U}(\underline{y-m}|qv+\underline{i-m}), \quad \forall \,
y,\,i\in\calS, \;v \in \calV,\label{eq:py|u-m}
\end{eqnarray}
where (\ref{eq:py|u-m-1}) follows from the definition
(\ref{eq:py|x-m}), and both (\ref{eq:py|u-m-2}) and
(\ref{eq:py|u-m-3}) follow by applying Lemma~\ref{lemma:pxu|s
property}(ii). We also obtain the marginal PMF of $Y$ as
\begin{eqnarray}
p_Y^m(y)&=&\sum_{v=0}^{L-1}\sum_{i=0}^{q-1}p_U(qv+i)p_{Y|U}^m(y|qv+i)\nonumber\\
&=&\sum_{v=0}^{L-1}\sum_{i=0}^{q-1}p_U(qv+\underline{i-m})p_{Y|U}(\underline{y-m}|qv+\underline{i-m})\label{eq:pym-1}\\
&=&p_Y(\underline{y-m}), \quad \forall\, y \in
\calS,\label{eq:pym}
\end{eqnarray}
where (\ref{eq:pym-1}) follows from Lemma~\ref{lemma:pxu|s
property}(iii) and (\ref{eq:py|u-m}).

From (\ref{eq:py|u-m}) and (\ref{eq:pym}), we obtain
\begin{equation}
I_{p_S, p_{XU|S}, p_{Y|X}}(Y;U)=I_{p_S, p_{XU|S}, p^m_{Y|X}}(Y;U)
\end{equation}
and hence
\begin{equation}
J_L(p_S, p_{XU|S}, p_{Y|X})=J_L(p_S, p_{XU|S}, p^m_{Y|X}),
\end{equation}
for $0 \le m<q$.

Let $\bar p_{Y|X} \triangleq \frac{1}{q}\sum_{m=0}^{q-1}
p^m_{Y|X}$. It is easy to check that $\bar p_{Y|X} \in
{\calA}_{cyc} (D_2)$. Also,
\begin{eqnarray}
J_L(p_S, p_{XU|S},p_{Y|X})&=&\frac{1}{q}\sum_{m=0}^{q-1} J_L(p_S,
p_{XU|S},p^m_{Y|X})\\
&\ge& J_L\left(p_S, p_{XU|S},\frac{1}{q}\sum_{m=0}^{q-1}
p^m_{Y|X}\right)=J_L(p_S, p_{XU|S},\bar p_{Y|X}), \label{eq:C
lower bound 1}
\end{eqnarray}
where the inequality comes from the fact that for fixed $p_S$ and
$ p_{XU|S}$, $J_L(p_S, p_{XU|S},p_{Y|X})$ is convex in
$p_{Y|X}$~\cite[Proposition 4.1(iii)]{Moulin_O'Sullivan_IEEE03}.
Therefore, from (\ref{eq:C lower bound 1}) we have
\begin{eqnarray}
C^{PubWM}(D_1,D_2)&\ge& C^{Steg}(D_1,D_2)\nonumber\\
&\ge& \lim_{L \to \infty}\max_{p_{XU|S} \in
\calQ_{cyc}^{Steg}(qL, p_S,D_1)} \min_{p_{Y|X} \in
\calA(p_X,D_2)}J_L(p_S,
p_{XU|S},p_{Y|X})\nonumber\\
&\ge&\lim_{L \to \infty}\max_{p_{XU|S} \in
\calQ_{cyc}^{Steg}(qL, p_S,D_1)} \min_{p_{Y|X} \in
{\calA}_{cyc}(D_2)}J_L(p_S, p_{XU|S},p_{Y|X}). \label{eq:C lower
bound}
\end{eqnarray}

Combining the upper bound inequality in (\ref{eq:C upper bound})
and the lower bound inequality in (\ref{eq:C lower bound}), we
prove the claim
\begin{eqnarray}
C^{PubWM}(D_1,D_2) &=&C^{Steg}(D_1,D_2)\nonumber\\
&=&\lim_{L \to \infty}\max_{p_{XU|S} \in
\calQ_{cyc}^{Steg}(qL, p_S,D_1)} \min_{p_{Y|X} \in
{\calA}_{cyc}(D_2)}J_L(p_S, p_{XU|S},p_{Y|X}),
\end{eqnarray}
which means that the perfectly secure steganographic constraint
does not cause any capacity loss.
\section{Proof of Lemma~\ref{lemma:Iyv<Iyu}}
\label{app:proof of lemma Iyv<Iyu}
For the pair $\left(p_{XV|S},p_{Y|X}\right) \in
 \calQ(L, p_S,D_1) \times \calA_{cyc}(D_2)$, the conditional PMF
of $Y$ given $V$ is
\begin{eqnarray}
p_{Y|V}(y|v)&=&\sum_{x=0}^{q-1}p_{X|V}(x|v)\,p_{Y|X}(y|x)\nonumber\\
&=&\sum_{x=0}^{q-1}p_{X|V}(x|v)\,p_{Y|X}(\underline{y-x}|\,0),
\quad \forall \; y \in \calS, \; v \in \calV, \label{eq:py|v}
\end{eqnarray}
where (\ref{eq:py|v}) follows from (\ref{eq:Acyc-py|x}) in
Definition~\ref{def:Acyc} for $p_{Y|X} \in \calA_{cyc}(D_2)$. The
conditional entropy of $Y$ given $V$ is
\begin{eqnarray}
H(Y|V)&=&-\sum_{v=0}^{L-1}p_V(v)\sum_{y=0}^{q-1}p_{Y|V}(y|v)\log
p_{Y|V}(y|v). \label{eq:Hy|v}
\end{eqnarray}

For the associated pair $\left(p_{XU|S},p_{Y|X}\right) \in
 \calQ_{cyc}^{Steg}(qL, p_S,D_1)\times \calA_{cyc}(D_2) $, the
conditional PMF of $Y$ given $U$ is
\begin{eqnarray}
p_{Y|U}\left(y|qv+i\right)&=&\sum_{x=0}^{q-1}p_{X|U}(x|qv+i)p_{Y|X}(y|x)\nonumber\\
&=&\sum_{x=0}^{q-1}p_{X|V}\left(\underline{x-i}\big|v\right)p_{Y|X}
	\left(\underline{y-i-(x-i)}\,\Big|\,0\right)\label{eq:py|u-1}\\
&=&p_{Y|V}(\underline{y-i}|v),\quad \forall \; y, \, i \in \calS,
\; v \in \calV, \label{eq:py|u}
\end{eqnarray}
where to obtain (\ref{eq:py|u-1}) we have used
Lemma~\ref{lemma:pxu|s property}(ii) and (\ref{eq:Acyc-py|x}) in
Definition~\ref{def:Acyc} for $p_{Y|X} \in \calA_{cyc}(D_2)$; and
(\ref{eq:py|u}) follows from (\ref{eq:py|v}). The marginal PMF of
$Y$ is given by
\begin{eqnarray}
\hat p_Y(y)&=&\sum_{v=0}^{L-1}\sum_{i=0}^{q-1}p_U(qv+i)\,p_{Y|U}(y|qv+i)\nonumber\\
&=&\sum_{v=0}^{L-1}\sum_{i=0}^{q-1}\frac{1}{q}p_V(v)\,p_{Y|V}(\underline{y-i}|v)\label{eq:py-1}\\
&=&\frac{1}{q}\sum_{j=0}^{q-1}
p_Y(\underline{j-i})=\frac{1}{q},\label{eq:py}
\end{eqnarray}
where (\ref{eq:py-1}) follows from Lemma~\ref{lemma:pxu|s
property}(iii) and (\ref{eq:py|u}). The conditional entropy of $Y$
given $U$ is
\begin{eqnarray}
H(Y|U)&=&-\sum_{v=0}^{L-1}\sum_{i=0}^{q-1}p_U(qv+i)\sum_{y=0}^{q-1}p_{Y|U}(y|qv+i)\,
\log p_{Y|U}(y|qv+i)\nonumber\\
&=&-\sum_{v=0}^{L-1}\sum_{i=0}^{q-1}\frac{1}{q}p_V(v)\sum_{y=0}^{q-1}
	p_{Y|V}(\underline{y-i}|v)\,\log p_{Y|V}(\underline{y-i}|v)\label{eq:Hy|u-1}\\
&=&\frac{1}{q}\sum_{j=0}^{q-1}H(Y|V)=H(Y|V),
\label{eq:Hy|u}
\end{eqnarray}
where (\ref{eq:Hy|u-1}) follows from Lemma~\ref{lemma:pxu|s
property}(iii) and (\ref{eq:py|u}), and (\ref{eq:Hy|u}) follows
from (\ref{eq:Hy|v}).

Since $\hat p_Y(y)=\frac{1}{q}$ for any $y \in \calS$ as shown in
(\ref{eq:py}), we have
\begin{equation}
H_{\hat p_Y}(Y)\ge H_{p_Y}(Y), \label{eq:Hy<hy}
\end{equation}
where $\hat p_Y$ and $p_Y$ are the marginal PMF of $Y$ for
$\left(p_S,p_{XU|S},p_{Y|X}\right)$ and
$\left(p_S,p_{XV|S},p_{Y|X}\right)$, respectively. Therefore, from
(\ref{eq:Hy|u}) and (\ref{eq:Hy<hy}), we obtain
\begin{eqnarray}
I(Y;U)&=&H_{\hat p_Y}(Y)-H(Y|U)\\
 &\ge& H_{p_Y}(Y)-H(Y|V)\\
&=&I(Y;V).\label{eq:Iyu}
\end{eqnarray}
Hence, Lemma~\ref{lemma:Iyv<Iyu} is proved.

\clearpage
\bibliographystyle{IEEEtran}
\bibliography{IEEEabrv,mybiblist}
\end{document}